\documentclass{aa}

\usepackage{verbatim}
\usepackage{graphicx}
\usepackage{epstopdf}
\usepackage{subfigure}
\usepackage{hyperref}
\usepackage{mathtools}
\usepackage{amsmath}

\title{Stellar activity and planetary atmosphere evolution in tight binary star systems\footnotemark}

\titlerunning{Activity and atmospheres in tight-binary star systems}

\author{C. P. Johnstone\inst{\ref{vienna}} \and E. Pilat-Lohinger\inst{\ref{vienna}} \and T. L\"{u}ftinger\inst{\ref{vienna}} \and M. G\"{u}del\inst{\ref{vienna}} \and A. St\"{o}kl\inst{\ref{vienna}}}

\institute{
University of Vienna, Department of Astrophysics, T\"{u}rkenschanzstrasse 17, 1180 Vienna, Austria \label{vienna}
}

\abstract{
In tight binary star systems, tidal interactions can significantly influence the rotational and orbital evolution of both stars, and therefore their activity evolution.
This can have strong effects on the atmospheric evolution of planets that are orbiting the two stars.
}{
In this paper, we aim to study the evolution of stellar rotation and of X-ray and ultraviolet (XUV) radiation in tight binary systems consisting of two solar mass stars and use our results to study planetary atmosphere evolution in the habitable zones of these systems.
}{
We have applied a rotation model developed for single stars to binary systems, taking into account the effects of tidal interactions on the rotational and orbital evolution of both stars.
We used empirical rotation-activity relations to predict XUV evolution tracks for the stars, which we used to model hydrodynamic escape of hydrogen dominated atmospheres. 
}{
When significant, tidal interactions increase the total amount of XUV energy emitted, and in the most extreme cases by up to factor of $\sim$50. 
We find that in the systems that we study, habitable zone planets with masses of 1~M$_\oplus$ can lose huge hydrogen atmospheres due to the extended high levels of XUV emission, and the time that is needed to lose these atmospheres depends on the binary orbital separation.
For some orbital separations, and when the stars are born as rapid rotators, it is also possible for tidal interactions to protect atmospheres from erosion by quickly spinning down the stars.
For very small orbital separations, the loss of orbital angular momentum by stellar winds causes the two stars to merge.
We suggest that the merging of the two stars could cause previously frozen planets to become habitable due to the habitable zone boundaries moving outwards.
}{}

\begin{document}

\maketitle

\footnotetext{Tabulated output data from all simulations used in this paper, accompanied by Python scripts used for making all figures, can be downloaded from \url{https://zenodo.org/record/2643479\#.XLcyO0NS9hE}. }


\section{Introduction}

In recent years, observational campaigns have found planets orbiting both components in binary star systems (e.g. \mbox{\citealt{2011Sci...333.1602D}}; \mbox{\citealt{2012Natur.481..475W}}) and it is likely that many tight binary systems possess planets on circumbinary orbits that are within the habitable zone.
The possibility of habitability in binary stars systems is very interesting, but the issue is often made more complicated by the presence of two stars (e.g. \citealt{Eggl13}).
When the binary orbital separation is small, tidal interactions between the two stars can influence their rotational evolution, and therefore their X-ray and extreme ultraviolet (together `XUV') evolution (\citealt{2014A&A...570A..50S}; \mbox{\citealt{2015IJAsB..14..391M}}; \citealt{Zuluaga16}).
This radiation is important for heating the upper atmospheres of planets and driving atmospheric mass loss (\mbox{\citealt{Lammer11}}).

A star's XUV emission depends primarily on its rotation rate, and therefore a description of its XUV evolution should be based on its rotational evolution. 
Single stars spin down as they age (\citealt{Skumanich72}), causing their activity levels to decay (\citealt{Guedel97}).
\mbox{\citet{Tu15}} showed that at young ages,  a star's XUV evolution depends strongly on its initial rotation rate, with stars that are born as fast rotators remaining highly active much longer than stars that are born as slow rotators.
In binary systems, the situation can be very different when the binary separation is small due to tidal interactions causing angular momentum exchange between the two components. 
In these cases, the additional tidal torque acts to synchronise the rotation periods of the two stars with their orbital period (\mbox{\citealt{1975A&A....41..329Z}}; \mbox{\citealt{1977A&A....57..383Z}}), influencing the rotational evolution of the stars (\mbox{\citealt{1997A&A...318..275K}}). 
When the two stars are strongly tidally interacting, they do not spin down as single stars, and they can remain active for much longer times. 
In such systems, the binary separation strongly influences their rotation rates and XUV emission levels.

Just how numerous tidally-locked binary systems are is an important question. 
\citet{Moe17} studied multiplicity in binary systems and found that approximately 30\% of solar-mass stars have binary companions, and a further 10\% are in triple or quadruple systems, with 15\% of solar-mass stars being close in systems, defined as systems with orbital separations less than 1~AU.
\citet{Lurie17} used rotation and orbital periods derived from Kepler data and found many systems that are tidally locked or very close to being tidally locked.
The survey of \mbox{\citet{2010ApJS..190....1R}} contained 454 systems, of which 44\% were found to be multiple systems. 
They found 11 systems (2.4\%) with orbital periods shorter than ten days, suggesting that a few percent of systems might contain tidally locked binaries.
Similar numbers can be inferred from the results of \mbox{\citet{2003A&A...397..159H}} and \mbox{\citet{Moe17}}.
Although tidally locked binaries make up a minority of systems, they are not negligible (a few percent of stellar systems corresponds to billions of systems in our galaxy) and are therefore interesting systems to study.

Observationally, there is clear evidence that short period ($\lesssim$10~days) binaries remain active. 
For example, the triple system $\kappa$~For consists of a pair of M-dwarfs that orbit each other with a 3.7~day period and a more distant Sun-like star.
Although the system is quite old, with the solar mass star being inactive, the two M dwarfs remain highly active, most likely due to tidal-locking (\mbox{\citealt{2013AJ....145...76T}}). 
Similarly, \citet{Frasca06} studied six close binary systems that showed evidence of tidal synchronisation between the orbits and rotation rates, all of which were highly active.
\citet{Dempsey93} and \citet{Dempsey97} studied X-ray activity in a large sample of binary systems of the RS CVn and BY Dra types.
All of the stars in their sample of RS CVn systems were rapid rotators with rotation periods lower than two days and all have high X-ray luminosities.
They concluded for RS CVn systems that the secondary has no effect on the X-ray activity levels of the stars, other than acting to tidally spin them up, and they found RS CVn and BY Dra systems do not differ significantly in X-ray properties.
\citet{Makarov03} study the most X-ray active stars in the solar neighbourhood and found that 40\% of the stars with X-ray luminosities above $10^{30}$~erg~s$^{-1}$ are short period binaries, which they explained as being due to them kept as rapid rotators by tidal interactions.
This interpretation is supported by the close correlation between X-ray luminosity and orbital period shown in Fig.~1 of \citet{Makarov09}, which they suggest is similar to the relation between X-rays and rotation for single stars.
Similarly, \mbox{\citet{2010ApJS..190....1R}} presented a survey of binary systems within 25~pc and showed that all systems with short orbital periods below 12~days are active (see their Fig.~18).

Another important effect of the tidal interactions in binary star systems is orbital decay over evolutionary timescales due to the transfer of orbital angular momentum to rotational angular momentum.
This takes place because stellar winds remove angular momentum from the two stars, and this angular momentum is replenished from the orbital angular momentum by tidal interactions.
In many cases, the orbits of the two stars decay significantly; this has been shown observationally to take place by several studies (\citealt{Eker06}; \citealt{2004MNRAS.349.1069K}).
These systems can eventually become contact binaries (\citealt{Jiang14}), many of which are known as W UMa systems.
\citet{2001A&A...370..157S} study X-ray activity in such systems and found that they were all very highly active, but were a factor of a few less active than the most active single stars, which they interpret as being a result of horizontal flows over the joined surfaces of the two stars reducing the surface filling factors of active regions. 

As they continue to lose angular momentum, contact binaries can coalesce to form a single star (\citealt{Andronov06}).
The amount of time that binaries spend in the contact phase is unclear, with estimates ranging from tens of Myr to several Gyr (\citealt{Chen08}).
A binary merger was observed to take place in the system V1309 Sco in 2008 (\citealt{Mason10}; \citealt{Tylenda11}).
Previous to 2008, the system showed periodic variations in its lightcurve with a period that was exponentially decreasing; in 2008, the brightness of the system increased by orders of magnitude and then decayed again over the course of approximately two years.

One of the most important effects of a star's XUV emission is its influence on planetary atmospheres. 
Such radiation is absorbed high in the atmosphere, which can lead to the upper atmosphere being heated to temperatures of >1000~K.
This heating can cause expansion of the atmosphere and mass loss (\mbox{\citealt{2005ApJ...621.1049T}}; \mbox{\citealt{2008A&A...483..933E}}). 
The mass loss rate depends strongly on the stellar XUV luminosity (e.g. \mbox{\citealt{2009ApJ...693...23M}}), which therefore implies that the rotational evolution of the host star is fundamentally important for the evolution of a planet's atmosphere.
Such a dependence was studied by \mbox{\citet{Johnstone15apj}}, who combined an atmospheric loss model (described later in this paper) with the XUV evolution tracks of \mbox{\citet{Tu15}} to show the importance of the initial rotation rate of the host star on the erosion of hydrogen dominated atmospheres.
In addition to thermal expansion driven mass loss, atmospheres can be eroded by non-thermal processes driven by interactions with the star's wind (\mbox{\citealt{2013AsBio..13.1030K}}; \mbox{\citealt{2014A&A...562A.116K}}).

Given the importance of stellar rotational evolution for planetary atmosphere evolution, it should be expected that in tidally interacting binary systems, the orbital separation is an important parameter for determining how an atmosphere evolves. 
The importance of these effects on planetary habitability was studied by \mbox{\citet{2013ApJ...774L..26M}} and \mbox{\citet{2015IJAsB..14..391M}} who found that tidal interactions can cause habitable-zone planets in such systems to receive less or more XUV radiation over their lifetimes. 
Another complication is the fact that the winds from both stars collide, producing shocks and regions of enhanced density and temperature which circumbinary planets must pass through multiple times per orbit (\mbox{\citealt{2015A&A...577A.122J}}).

In this paper, we study the evolution of hydrogen dominated atmospheres in tidally interacting binary systems.
We concentrate on the case of two solar mass stars on circular orbits; in addition, we assume the two stars are being orbited by an Earth mass terrestrial planet at 1.5~AU, such that it has an effective temperature of $\sim$250~K. 
The stability for a planet at 1.5~AU is given for distances of the two G-type stars up to 0.66~AU according to Fig.~1 in \citet{PilatLohinger2003}. 
Moreover, the planet's orbit is in the habitable zone (HZ) of the binary system which extends from 1.34 to 2.37~AU, as calculated using the method of \citet{Kopparapu2014} for single stars.
In view of the fact that many stars are in tidally interacting binary systems, such binary-star--planet systems are interesting configurations also for the search of habitable planets. 
We assume that the planet collected a hydrogen dominated protoatmosphere during the disk phase of the system and concentrate on the evolution of this atmosphere. 

In Section~\ref{sect:rotorbevo}, we discuss rotational evolution in tidally interacting binaries.
In Section~\ref{sect:XUVevo}, we study the resulting XUV evolution in these systems.
In Section~\ref{sect:atmosphereevo}, we model the atmospheric evolution in our assumed system.
In Section~\ref{sect:conclusions}, we discuss our results and their significance.

\section{Rotational and orbital evolution in tight binary systems} \label{sect:rotorbevo}

\subsection{Rotation and orbit evolution model} \label{sect:rotmodel}

The rotational and orbital evolution of the stars in a tight binary system can be described just as the rotational evolution of two single stars with the addition of tidal interactions. 
How the rotation rates of solar mass stars evolve is observationally mostly well understood, primarily from observations of rotation rate distributions in young stellar clusters.
However, we currently lack a complete understanding of the physical mechanisms involved, although likely all of the most important mechanisms have been identified.
Rotational evolution models currently require the use of several free parameters that are tuned to fit the observational constraints.
For a comprehensive review of the topic, see \citet{Bouvier14}.

At ages of $\sim$1~Myr, stars of a given mass have a very broad distribution of rotation rates. 
Since they are contracting, pre-main-sequence (PMS) stars should spin up as they age. 
For the first few Myr, however, this appears not to be the case, and instead it seems that this distribution does not evolve significantly; this phenomenon is called `disk locking' and is poorly understood (\citealt{Matt10}).
At about the ages that circumstellar gas disks disappear, stars start to spin-up as expected.
During this phase, the spin-up is being counteracted by the spin-down from the removal of angular momentum by stellar winds, which we expect are much stronger than the current solar wind. 
As stars reach the zero-age main-sequence (ZAMS), their contractions slow and then stop, and the stellar wind driven spin-down takes over.
At this point, the initially broad distribution of rotation rates has become even broader, possible due to rapidly rotating stars spending less time in the disk-locking phase (\citealt{2013A&A...556A..36G}). 
Over the course of the following Gyr, the fastest rotators spin down rapidly and the distribution of rotation rates converge; however, these stars spin-down much slower than might be expected due to the saturation of magnetic activity at high rotation rates.
Rotational evolution models have had trouble accurately reproducing distributions of rotation rates in young stellar clusters during this phase without assuming that the stars do not rotate as solid bodies.
Instead, they must assume that the inner regions of the stars rotating faster at the ZAMS than the surfaces (\citealt{Krishnamurthi97}).
If angular momentum transport within stars takes place on timescales of >10~Myr, this would happen since during the PMS spin-up phase, stellar winds would only be directly spinning down the outer regions of stars. 

We use the rotational evolution model for single stars presented in \mbox{\citet{Tu15}}, which is based on the model of \mbox{\citet{2015A&A...577A..28J}} and is very similar to the model described in \mbox{\citet{2013A&A...556A..36G}}.
For the evolutions of all parameters related to the star's internal structure, including total stellar radius, we use the stellar evolution models of \mbox{\citet{Spada13}} for a solar mass star with an initial metallicity, composition, and mixing length parameter most closely matching the values for the Sun.
Their results include estimates for the time evolution of the stellar convective turnover time.
In our rotational evolution model, we assume each star is composed of a core and an envelope, and each of these components is assumed to be rotating as a solid body with its own rotation rate.
The envelope is the outer convective zone and the core is everything interior to this. 
In addition, we evolve the binary orbit, meaning that we have in total five reservoirs of angular momentum in our system and five quantities to evolve: these five quantities are core and envelope rotation rates, $\Omega_\mathrm{core}$ and $\Omega_\mathrm{env}$, for both stars, and the binary orbital separation $a_\mathrm{orb}$.
In this section, we give the system of ordinary differential equations (ODEs) that we use to evolve these five quantities.
From this point in the text until we describe the evolution of $a_\mathrm{orb}$, we describe the rotation model as it is applied to both stars individually and each of the equations is applied to both stars separately.

The angular momentum of a rotating body is \mbox{$J = I \Omega$}, where $I$ and $\Omega$ are the moment of inertia and angular velocity. 
Since all three of these quantities can be changing with time, the rates of change are related by
\begin{equation} \label{eqn:dJdt}
\tau = \frac{dJ}{dt} = I \frac{d\Omega}{dt} + \Omega \frac{dI}{dt},
\end{equation}
where $\tau$ is the torque.
Applying this equation to the core and envelope of a star gives
\begin{equation} \label{eqn:dOmegaCoredt}
\frac{d\Omega_\mathrm{core}}{dt} = \frac{1}{I_\mathrm{core}} \left( -\tau_\mathrm{ce} - \tau_\mathrm{cg} - \Omega_\mathrm{core} \frac{d I_\mathrm{core} }{dt} \right),
\end{equation}
\begin{equation} \label{eqn:dOmegaEnvdt}
\frac{d\Omega_\mathrm{env}}{dt} = \frac{1}{I_\mathrm{env}} \left( \tau_\mathrm{w} + \tau_\mathrm{ce} + \tau_\mathrm{cg} + \tau_\mathrm{dl} + \tau_\mathrm{ts} - \Omega_\mathrm{env} \frac{d I_\mathrm{env} }{dt} \right).
\end{equation}
Here, we have replaced the torque, $\tau$, with the sums of all the torques acting on the two zones. 
These are the stellar wind spin-down torque, $\tau_\mathrm{w}$, the core-envelope coupling torque, $\tau_\mathrm{ce}$, the core-growth torque, $\tau_\mathrm{cg}$, the disk-locking torque, $\tau_\mathrm{dl}$, and the tidal synchronisation torque, $\tau_\mathrm{ts}$.

Stellar winds remove mass from the surfaces of stars, and because they are fully ionised and therefore coupled to the star's magnetic field, they remove significant amounts of angular momentum (\citealt{WeberDavis67}).
The angular momentum is transported away from the star in two forms: as the angular momentum of the material itself, and as stresses in the magnetic field, with the latter form dominating close to the star (\citealt{Vidotto14}). 
The spin down torque, $\tau_\mathrm{w}$, acts directly on the envelope and is in our model always negative by definition. 
We calculate $\tau_\mathrm{w}$ as 
\begin{equation} \label{eqn:modifyMatttorque}
\tau_\mathrm{w} = - K_\tau \tau',
\end{equation}
where \mbox{$K_\tau=11$} is a free parameter derived by \mbox{\citet{2015A&A...577A..28J}} based on considering the Sun.
For $\tau'$, we use
\begin{equation} \label{eqn:matttorque}
\tau' = K_1^2 B_\mathrm{dip}^{4m} \dot{M}_\star^{1-2m} R_\star^{4m+2} \frac{\Omega_\mathrm{env}}{(K_2^2 v_\mathrm{esc}^2 + \Omega_\mathrm{env}^2 R_\star^2)^m},
\end{equation}
where $B_\mathrm{dip}$ is the strength of the dipole component of the star's magnetic field, $\dot{M}_\star$ is the wind mass loss rate, $R_\star$ is the stellar radius, $M_\star$ is the stellar mass, and $v_\mathrm{esc}$ is the surface escape velocity (\mbox{$=\sqrt{2 G M_\star / R_\star}$}).
This equation was derived by \mbox{\citet{2012ApJ...754L..26M}} using a grid of 2D magnetohydrodynamic wind simulations and assuming that the stellar magnetic fields are dipolar; they found $K_1 = 1.3$, $K_2 = 0.0506$, and $m = 0.2177$. 
Several recent studies have shown that details of the stellar magnetic field structure and the heating and acceleration mechanisms influence the rates at which stars lose angular momentum (e.g. \citealt{Reville15}; \citealt{Garraffo16}; \citealt{Cohen17}; \citealt{Pantolmos17}).
For example, if the field is not fully dipolar, the rate at which angular momentum is lost will have a different dependence on field strength than what is given in the above equation, though it is likely that the dipole component still dominates (\citealt{FinleyMatt18}).
For our purposes in this paper, it is not necessary to consider these details since we only need equations that give a reasonable description of single star rotational evolution.

The dipole field and the mass loss are both manifestations of the star's magnetic activity, and are both expected to depend sensitively on the star's rotation rate.
For slow rotators, we expect that there are power law dependences of $\dot{M}_\star$ and $B_\mathrm{dip}$ on the rotation rate, and for fast rotators, magnetic activity saturates, such that the rotation dependence disappears. 
However, on the pre-main-sequence (<40~Myr for solar mass stars), age should also be a factor, such that very young stars are saturated even at slow rotation.
This can be reproduced if we describe the dependence of magnetic activity on rotation using the Rossby number, $Ro_\star$, defined as \mbox{$P_\mathrm{rot}/t_\mathrm{conv}$}, where $t_\mathrm{conv}$ is the convective turnover time.  
We assume that \mbox{$Ro_\mathrm{sat}=0.13$} is the Rossby number separating the two regimes (i.e. when $Ro_\star < Ro_\mathrm{sat}$, stars are saturated).
This is based on the value derived from stellar X-ray emission by \mbox{\citet{Wright11}}. 
For the dipole field strength, we use 
\begin{equation} \label{eqn:dipoleRossby}
B_\mathrm{dip} = \left \{
\begin{array}{ll}
B_{\mathrm{dip},\odot} \left( \frac{Ro_\star}{Ro_\odot} \right)^{-1.32}, & \text{if }  Ro_\star \ge Ro_\mathrm{sat},\\
B_{\mathrm{dip},\odot} \left( \frac{Ro_\mathrm{sat}}{Ro_\odot} \right)^{-1.32}, & \text{if }  Ro_\star \le Ro_\mathrm{sat},\\
\end{array} \right.
\end{equation}
where $Ro_\odot$ is the solar Rossby number, and \mbox{$B_{\mathrm{dip},\odot} = 1.35$~G} (\mbox{\citealt{2015A&A...577A..28J}}).  
The index of -1.32 was derived by \mbox{\citet{2014MNRAS.441.2361V}} based on magnetic field measurements of a large sample of stars. 
As in \mbox{\citet{Tu15}}, we assume the mass loss rate is given by
\begin{equation} \label{eqn:MdotRossby}
\dot{M}_\star = \left \{
\begin{array}{ll}
\dot{M}_\odot \left( \frac{R_\star}{R_\odot} \right)^2 \left( \frac{Ro_\star}{Ro_\odot} \right)^{-2}, & \text{if }  Ro_\star \ge Ro_\mathrm{sat},\\
\dot{M}_\odot \left( \frac{R_\star}{R_\odot} \right)^2 \left( \frac{Ro_\mathrm{sat}}{Ro_\odot} \right)^{-2}, & \text{if }  Ro_\star \le Ro_\mathrm{sat},\\
\end{array} \right.
\end{equation}
where \mbox{$\dot{M}_\odot = 1.4 \times 10^{-14}$~M$_\odot$~yr$^{-1}$} is the current Sun's mass loss rate.
This is a modified version of the formula given by \mbox{\citet{2015A&A...577A..28J}}, who found \mbox{$\dot{M}_\star \propto R_\star^2 \Omega_\star^{1.33} M_\star^{-3.36}$} based on consideration of main-sequence stellar evolution.
The two expressions are different because we only consider solar mass stars here, so the $M_\star$ dependence is not necessary, and because we consider also the pre-main-sequence rotational evolution, meaning that Rossby number is a better quantity to use than $\Omega_\star$, as explained above.
\mbox{\citet{2015A&A...577A..28J}} also likely underestimated the dependence of $\dot{M}_\star$ on $\Omega_\star$ because they did not consider core-envelope decoupling in their model; models that do consider core-envelope decoupling require larger wind torques to spin-down the stars on the early main-sequence because the stars contain more angular momentum than would be expected based simply on the surface rotation rates.

In the simplified two zone model for stellar internal rotation, the envelope (the outer convective zone) and the core (everything else) must exchange angular momentum at the boundary between the two (the tachocline). 
It is not well understood how angular momentum is transported within stars, and it is not possible for us to include a detailed physical description of this angular momentum exchange. 
For the core-envelope coupling torque, $\tau_\mathrm{ce}$, we use the simplified model described in \citet{MacGregorBrenner91} and \mbox{\citet{2015A&A...577A..98G}}, where the torque is given by
\begin{equation} \label{eqn:CEtorque}
\tau_\mathrm{ce} = \frac{\Delta J}{ t_\mathrm{ce} },
\end{equation}
where $t_\mathrm{ce}$ is the core-envelope coupling timescale and \mbox{$\Delta J$} is the angular momentum that at a given time would need to be transferred between the two components in order to make them rotate with the same speed.
Assuming that a positive torque means that angular momentum is being taken from the core and given to the envelope, as implied in Eqns.~\ref{eqn:dOmegaCoredt} and \ref{eqn:dOmegaEnvdt}, the latter is given by
\begin{equation} \label{eqn:deltaJ}
\Delta J 
= 
\frac{ I_\mathrm{env} I_\mathrm{core} }{ I_\mathrm{env} + I_\mathrm{core} } 
\left( \Omega_\mathrm{core} - \Omega_\mathrm{env} \right).
\end{equation}
When \mbox{$\Omega_\mathrm{core}=\Omega_\mathrm{env}$}, this gives \mbox{$\Delta J=0$}, and therefore we get no angular momentum exchange between the core and the envelope.
When \mbox{$\Omega_\mathrm{core} > \Omega_\mathrm{env}$}, $\tau_\mathrm{ce}$ is positive, which means there is a spin-down torque exerted on the core and a spin-up torque exerted on the envelope.
In this model, we assume $t_\mathrm{ce}$ has a power-law dependence on the difference between the envelope and core rotation rates, given by
\begin{equation} \label{eqn:tCE}
t_\mathrm{ce} = a_\mathrm{ce} (|\Omega_\mathrm{env} - \Omega_\mathrm{core}|)^{b_\mathrm{ce}},
\end{equation}
where $t_\mathrm{ce}$ is in Myr and both $\Omega_\mathrm{env}$ and $\Omega_\mathrm{core}$ are in $\Omega_\odot$.
A similar assumption was made by \citet{2011MNRAS.416..447S}.
We have run a grid of rotational evolution models for single stars with a solar mass, and with different values of the free parameters, and picked by eye which give the best fits to the observational constraints given in \mbox{\citet{2015A&A...577A..28J}} and \mbox{\citet{Tu15}}.
We find \mbox{$a_\mathrm{ce} = 30.0$} and \mbox{$b_\mathrm{ce} = -0.2$}.

In addition to the core-envelope coupling torque, angular momentum is exchanged between the two zones in another way. 
As the core grows on the PMS, its mass and radius increase because material that is part of the envelope at the edge of the core-envelope boundary becomes part of the core.
As this material becomes part of the core, the angular momentum of the core increases and the angular momentum of the envelope decreases. 
We call this the core-growth torque, and it is given by
\begin{equation} \label{eqn:CGtorque}
\tau_\mathrm{cg} = - \frac{2}{3} R_\mathrm{core}^2 \Omega_\mathrm{env} \frac{dM_\mathrm{core}}{dt},
\end{equation}
where $R_\mathrm{core}$ and $M_\mathrm{core}$ are the core radius and mass. 
The minus sign in the above formula means that as the core is growing, the envelope is losing angular momentum and the core is gaining it.
The core-growth torque balances the final terms in Eqns.~\ref{eqn:dOmegaCoredt} and \ref{eqn:dOmegaEnvdt} involving $dI_\mathrm{env}/dt$ and $dI_\mathrm{core}/dt$; as the cores grow, their moments of inertia increase, but they don't spin down because they are also gaining angular momentum.
We note that the above is only valid when the core is growing, and therefore \mbox{$dM_\mathrm{core}/dt > 0$}; if the core was instead shrinking, with \mbox{$dM_\mathrm{core}/dt < 0$}, the $\Omega_\mathrm{env}$ term should be replaced by $\Omega_\mathrm{core}$.

Another important ingredient is disk-locking. 
Observationally, it is known that the distributions of rotation rates for stars is approximately constant in the first few Myr when they still possess circumstellar gas disks, despite the fact that they are contracting and therefore should be spinning up (\citealt{Rebull04}).
Disk-locking is thought to be a result of interactions between the star and the disk, though it is surprising that these interactions would remove angular momentum from the star given that the accretion of disk material onto the stellar surface should increase the star's specific angular momentum. 
It is not known what causes `disk-locking' and many processes have been proposed (for a review, see \citealt{Bouvier14}); one promising idea is that accretion onto the surface enhances the star's wind, leading to enhanced wind driven spin-down (\citealt{MattPudritz08}; \citealt{Cranmer09}).
As is normal in rotational evolution models, \mbox{\citet{Tu15}} modelled disk-locking simply by setting \mbox{$d\Omega_\mathrm{env}/dt=0$} at ages less than the disk locking time, $t_\mathrm{disk}$.
However, for our purposes, this is likely inappropriate: when the binary separation is very small, it is not reasonable to assume that the disk-locking torque is able to overcome the tidal synchronisation torque.
We therefore include in Eqn.~\ref{eqn:dOmegaEnvdt} a disk-locking torque given by
\begin{equation} \label{eqn:disklocking}
\tau_\mathrm{dl} = \left \{
\begin{array}{ll}
-\tau_\mathrm{w} - \tau_\mathrm{ce} - \tau_\mathrm{cg} + \Omega_\mathrm{env} \frac{d I_\mathrm{env} }{dt}, & \text{if } t \le t_\mathrm{disk},\\
0, & \text{otherwise}.\\
\end{array} \right.
\end{equation}
This has the effect of canceling out all terms in Eqn.~\ref{eqn:dOmegaEnvdt} except the tidal synchronisation torque. 
When the tidal synchronisation torque is negligible, disk-locking takes place, and when it is dominant, tidal synchronisation takes place.
For the disk-locking time, we use the simple scaling law given by \mbox{\citet{Tu15}} of 
\begin{equation} \label{eqn:timeCE}
t_\mathrm{disk} = 13.5 \Omega_0^{-0.5},
\end{equation}
where $\Omega_0$ is the initial rotation rate of the star in units of $\Omega_\odot$ and $t_\mathrm{disk}$ is in Myr. 
This means that the envelopes of fast rotators start to spin up earlier as required to reproduce the observed fast rotators in the young $\sim$13~Myr old cluster h~Per (\mbox{\citealt{2013A&A...560A..13M}}).

The final torque that we need in our model is the tidal synchronisation torque.
For zero eccentricity, the tidal synchronisation torque can be calculated using
\begin{equation} \label{eqn:tauTS}
\tau_\mathrm{ts} = \frac{J_\mathrm{orb}}{2 T_\star} \left( 1 - \frac{\Omega_\mathrm{env}}{\Omega_\mathrm{orb}} \right),
\end{equation}
where $J_\mathrm{orb}$ and $\Omega_\mathrm{orb}$ are the orbital angular momentum and angular velocity, and $T_\star$ is the dissipation timescale (see for example Eqn.~5 of \mbox{\citealt{2012A&A...544A.124B}}).
To calculate the dissipation timescale, we consider the effects of both equilibrium tides and dynamical tides, with the difference in the two mechanisms in our model being based on how we calculate $T_\star$.
As is common in the literature, we assume dynamical tides dominate when \mbox{$\omega \in [ -2\Omega_\star , 2 \Omega_\star ]$}, where $\omega$ is the excitation frequency given by \mbox{$\omega = 2 ( \Omega_\mathrm{orb} - \Omega_\star )$}, and equilibrium tides dominate otherwise.

For equilibrium tides, we use the model developed by \citet{Eggleton98} in the form given by \citet{2012A&A...544A.124B}.
In this model, we calculate the dissipation timescale using
\begin{equation} \label{eqn:disstimeEQ}
T_\star = \frac{ M_\star a_\mathrm{orb}^8 }{ 9 M_2 \left( M_\star + M_2 \right) R_\star^{10} \sigma_\star },
\end{equation}
where $M_2$ is the mass of the companion, $a_\mathrm{orb}$ is the orbital separation, and $\sigma_\star$ is the tidal dissipation factor, which we set to \mbox{$4.992 \times 10^{-66}$~g$^{-1}$~cm$^{-2}$~s$^{-1}$} as estimated empirically by \mbox{\citet{2010ApJ...723..285H}}.
It is important that the tidal synchronisation torque depends sensitively on the stellar radius.

For dynamical tides, we use the model used by \citet{Gallet18} and developed by \citet{Ogilvie13}.
The dissipation timescale is calculated using
\begin{equation} \label{eqn:disstimeDY}
T_\star = \frac{ 2 M_\star a_\mathrm{orb}^8 \bar{Q_\mathrm{s}'} | \Omega_\mathrm{orb} - \Omega_\mathrm{env} | }{ 9 G M_2 \left( M_\star + M_2 \right) R_\star^{5} \hat{\epsilon}^2 },
\end{equation}
where \mbox{$\hat{\epsilon} = \Omega_\mathrm{env} / \sqrt{G M_\odot / R_\odot^3}$} and $\bar{Q_\mathrm{s}'}$ is the equivalent modified tidal quality factor given by \mbox{$\bar{Q_\mathrm{s}'} = 3 \Omega_\mathrm{env}^2/(2 \Omega_\mathrm{c}^2 \langle D \rangle_\omega )$}, where \mbox{$\Omega_\mathrm{c}=G M_\star / R_\star^3$} is the critical angular velocity and $\langle \mathcal{D} \rangle$ is given by
\begin{equation}
\begin{split}
\langle \mathcal{D} \rangle_\omega & = \frac{100 \pi}{63} \epsilon^2 \left( \frac{\alpha^5}{1-\alpha^5} \right) \left( 1 - \gamma \right)^2 \left( 1 - \alpha \right)^4 \times \\
& \left( 1 + 2 \alpha + 3 \alpha^2 + \frac{3}{2} \alpha^3 \right)^2 \left[ 1 + \left( \frac{1-\gamma}{\gamma} \right) \alpha^3 \right] \times \\
& \left[ 1 + \frac{3}{2} \gamma + \frac{5}{2\gamma} \left( 1 + \frac{1}{2} \gamma - \frac{3}{2} \gamma^2 \right) \alpha^3 - \frac{9 \left(1-\gamma \right)}{4}  \alpha^5 \right]^{-2} ,
\end{split}
\end{equation}
where
\begin{equation}
 \alpha = \frac{R_\mathrm{rad}}{R_\star}, \quad  \beta = \frac{M_\mathrm{rad}}{M_\star}, \quad \gamma = \frac{\alpha^3 \left( 1 - \beta \right)}{ \beta \left( 1 - \alpha^3 \right)} , \quad \epsilon = \frac{\Omega_\mathrm{env}}{\Omega_\mathrm{c}}.
\end{equation}

\begin{figure}
\centering
\includegraphics[trim = 0mm 0mm 0mm 0mm, clip=true,width=0.49\textwidth]{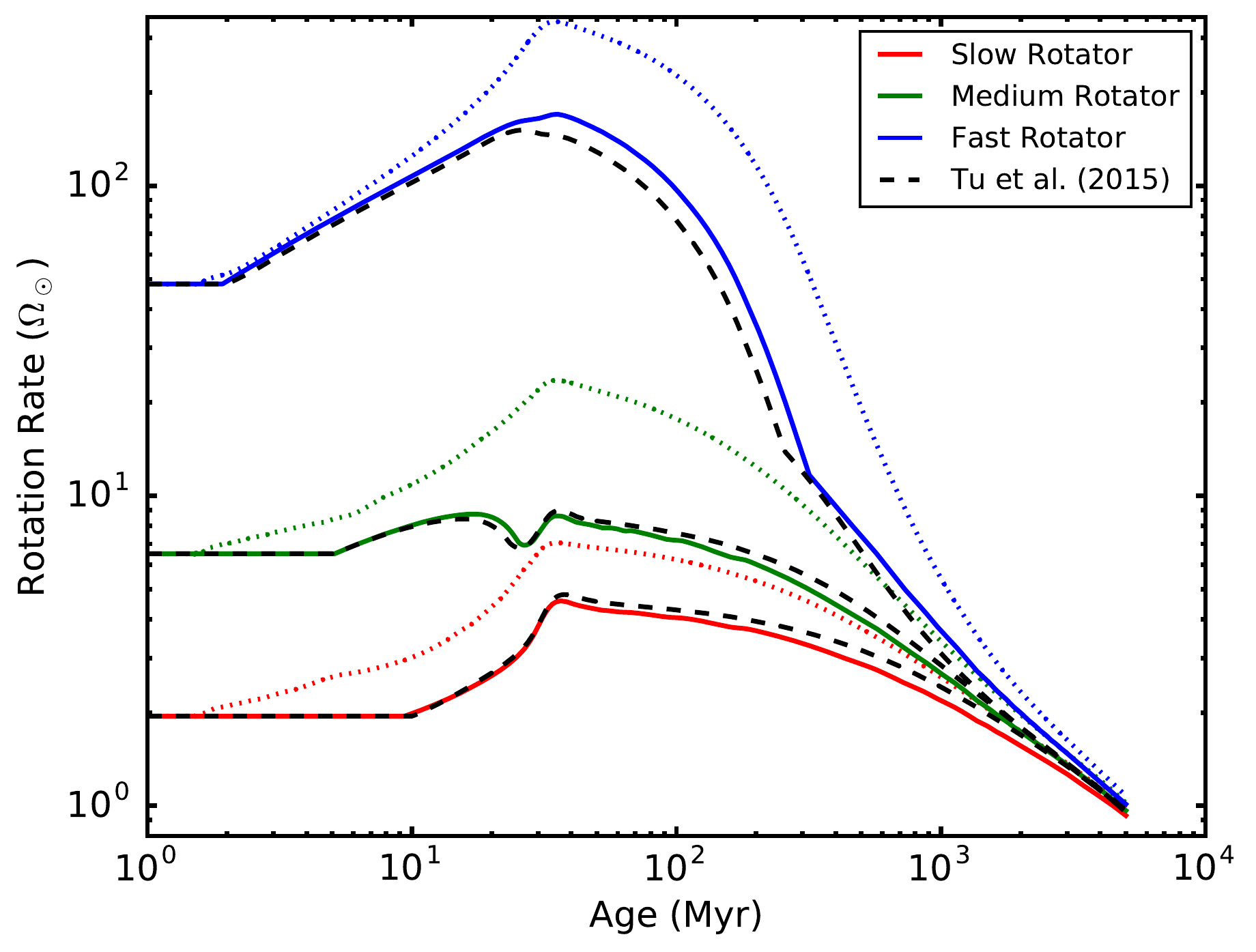}
\caption{
Rotational evolution of the envelope (solid lines) and core (dotted lines) for three solar mass stars with different initial rotation rates.
The black dashed lines are the models of \mbox{\citet{Tu15}}, which we use for comparison.
}
 \label{fig:solarrotation}
\end{figure}

The final ingredient in our evolutionary model is the time evolution of the orbital separation.
The orbital angular momentum of a binary system with zero eccentricity is
\begin{equation} \label{eqn:Jorb}
J_\mathrm{orb} =  \frac{M_1 M_2}{\left( M_1 + M_2 \right)^{1/2}}  G^{1/2}  a_\mathrm{orb}^{1/2},
\end{equation}
where the sub-scripts 1 and 2 indicate the quantities for the two stars.
The angular momentum lost by the orbit is given by
\begin{equation} \label{eqn:dJorbdt}
\frac{dJ_\mathrm{orb}}{dt} =  - \left( \tau_\mathrm{ts,1} + \tau_\mathrm{ts,2} \right),
\end{equation}
where $\tau_\mathrm{ts,1}$ and $\tau_\mathrm{ts,2}$ are the tidal synchronisation torques for the two stars.
The minus sign in the above equation is there because we define the tidal synchronisation torque such that positive values mean that the star is gaining angular momentum and the orbit is losing it.
Differentiating Eqn.~\ref{eqn:Jorb} with respect to time therefore gives the following equation for the orbital evolution
\begin{equation} \label{eqn:daorbdt}
\frac{da_\mathrm{orb}}{dt} = -2 \frac{ \left( M_1 + M_2 \right)^\frac{1}{2} }{ M_1 M_2 G^\frac{1}{2} } \left( \tau_\mathrm{ts,1} + \tau_\mathrm{ts,2} \right)
 a_\mathrm{orb}^\frac{1}{2}.
\end{equation}
If too much of the orbital angular momentum is removed, it is possible that the two stars merge.
We do not treat the detailed physics of stellar mergers in this paper, but instead stop our models if the orbital separation becomes smaller than the sum of the undisturbed radii of the two stars.

\begin{figure*}
\centering
\includegraphics[trim = 0mm 0mm 0mm 0mm, clip=true,width=0.42\textwidth]{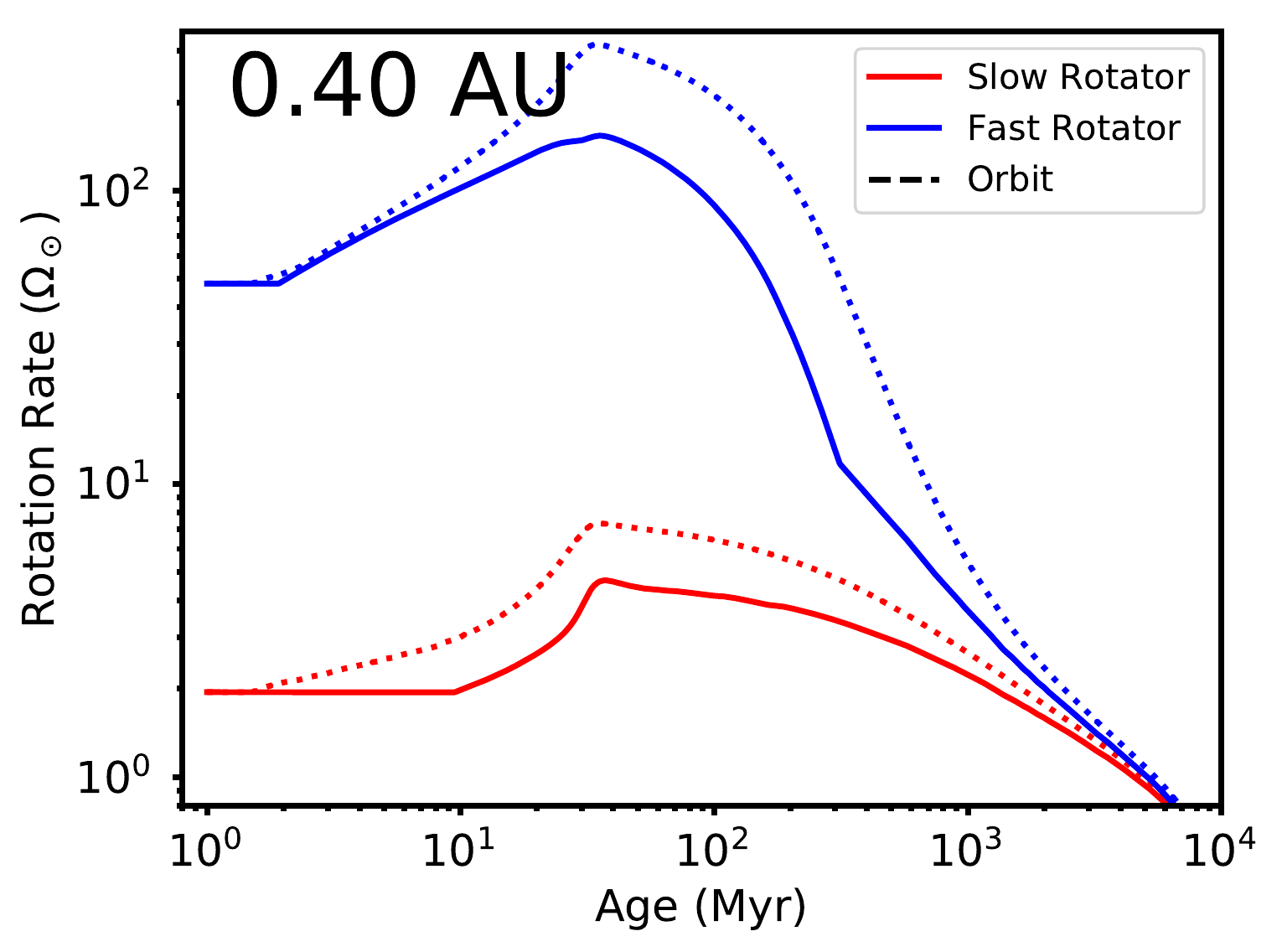}
\includegraphics[trim = 0mm 0mm 0mm 0mm, clip=true,width=0.42\textwidth]{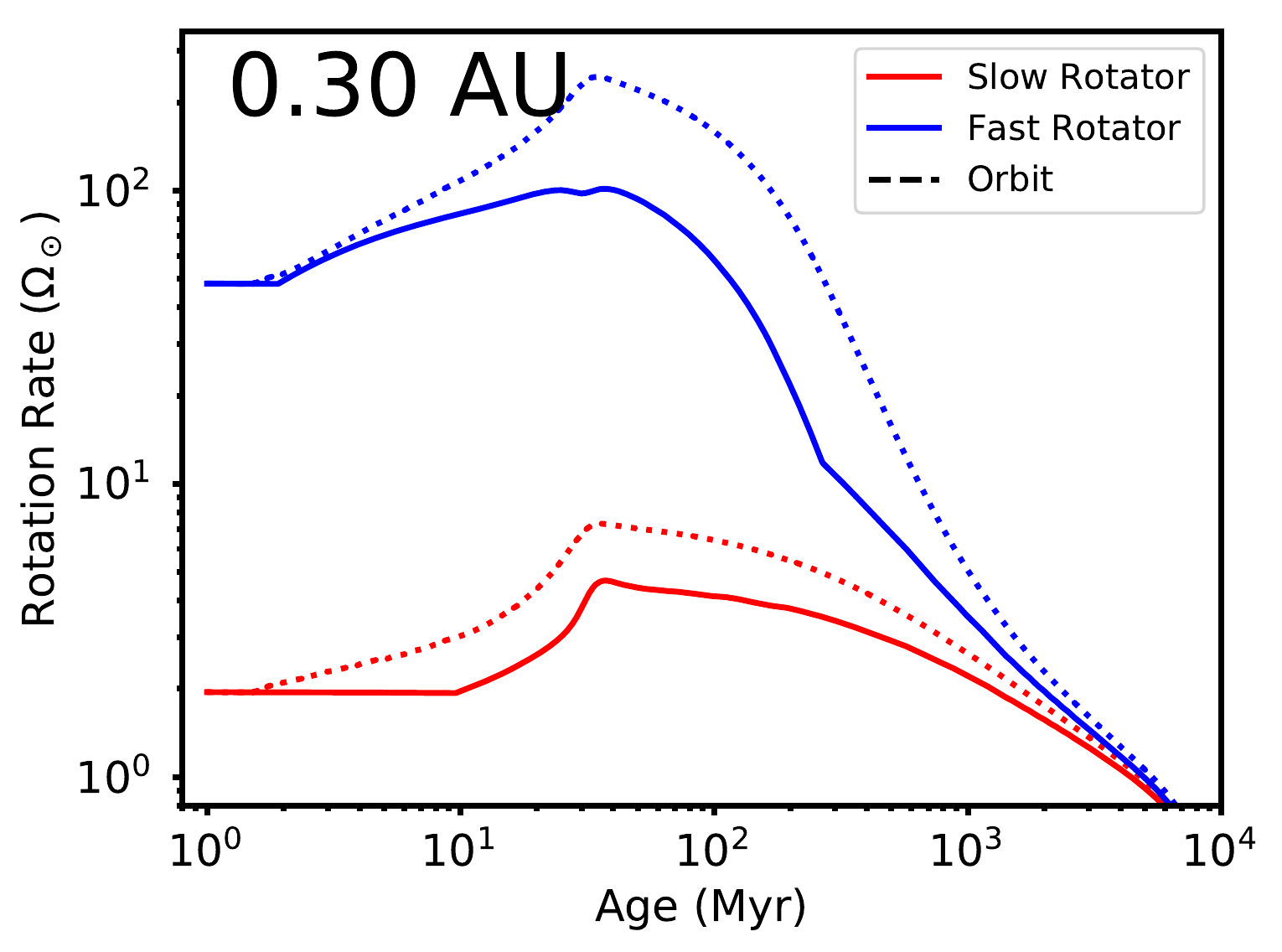}
\includegraphics[trim = 0mm 0mm 0mm 0mm, clip=true,width=0.42\textwidth]{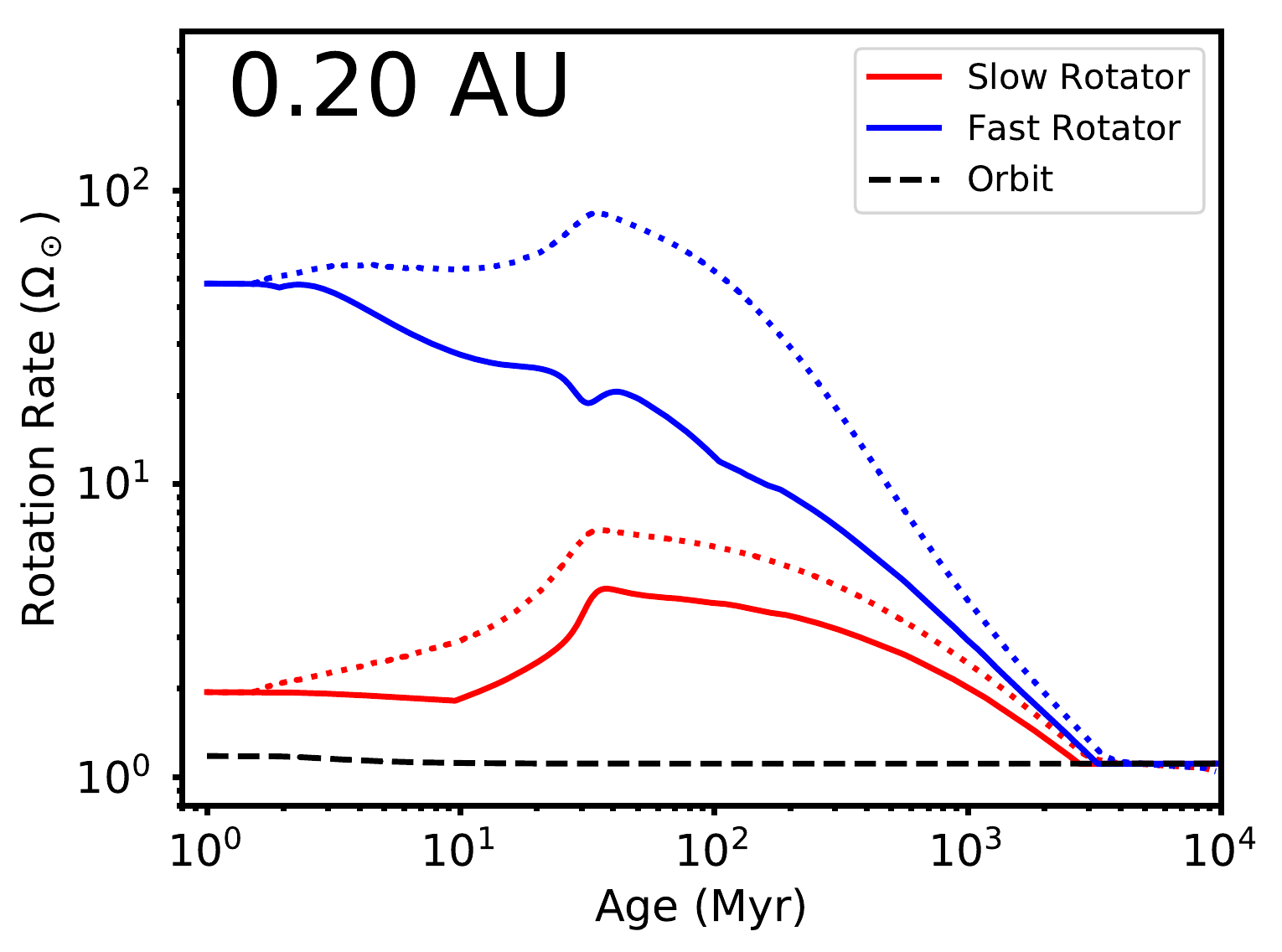}
\includegraphics[trim = 0mm 0mm 0mm 0mm, clip=true,width=0.42\textwidth]{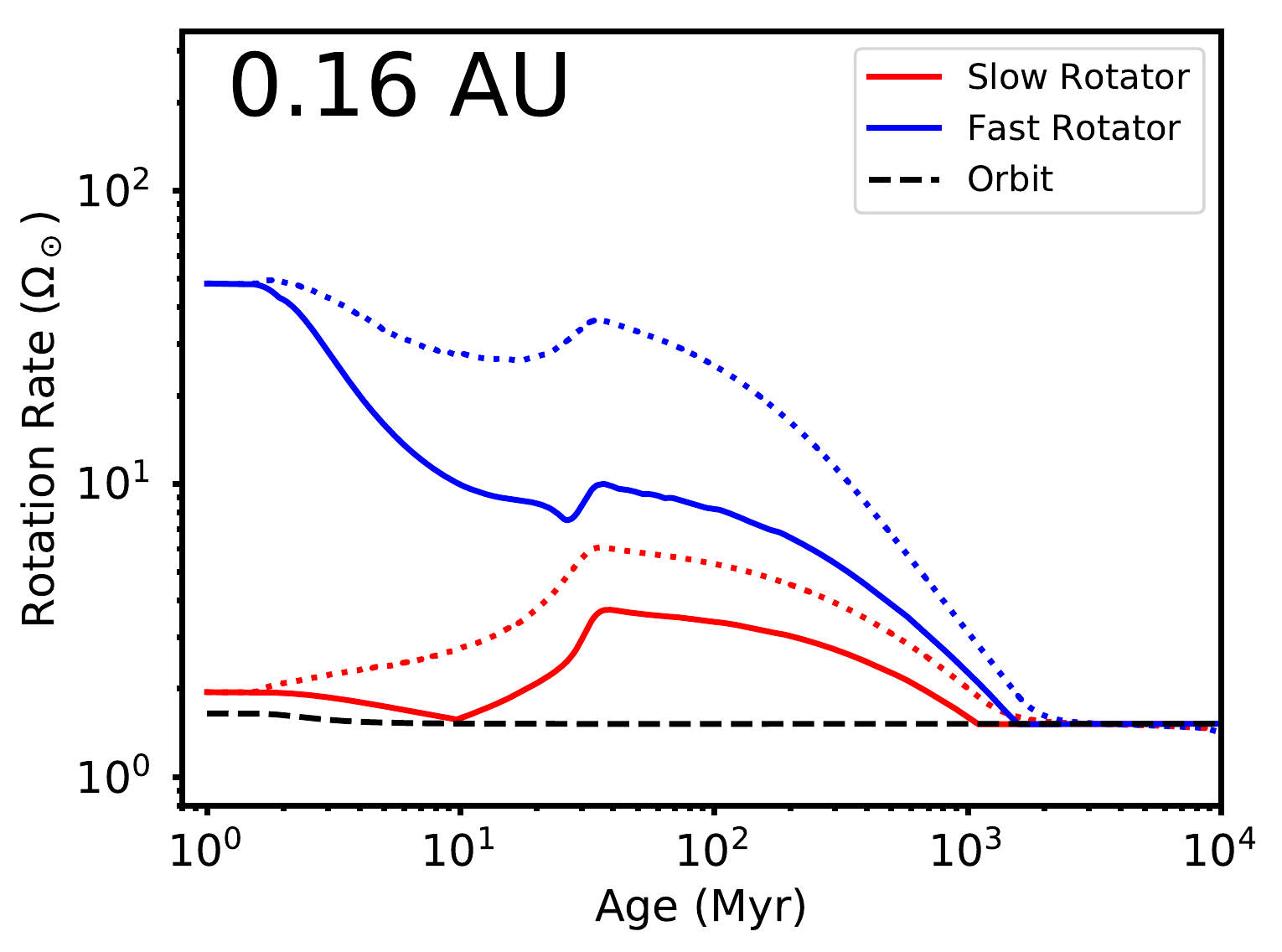}
\includegraphics[trim = 0mm 0mm 0mm 0mm, clip=true,width=0.42\textwidth]{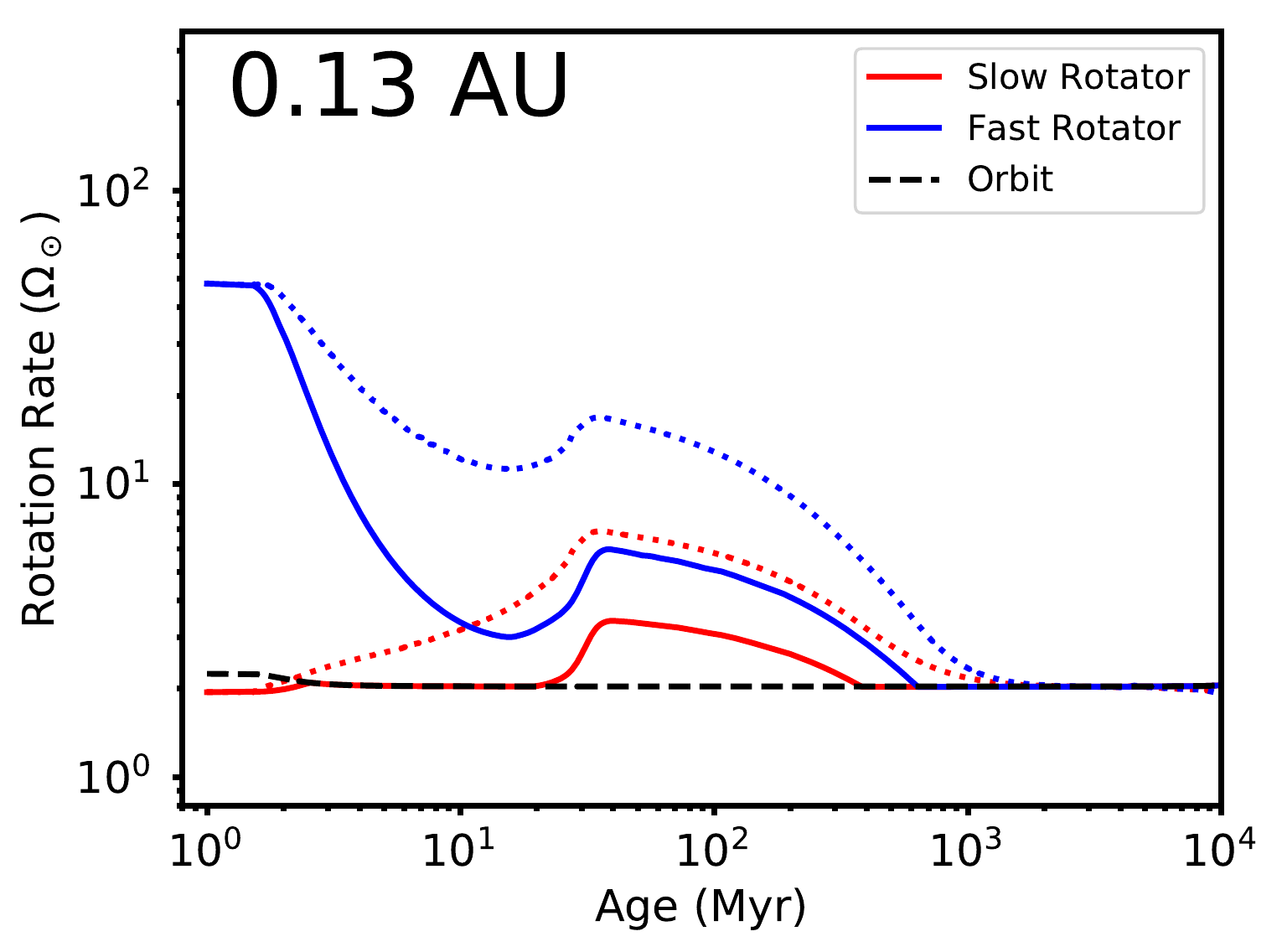}
\includegraphics[trim = 0mm 0mm 0mm 0mm, clip=true,width=0.42\textwidth]{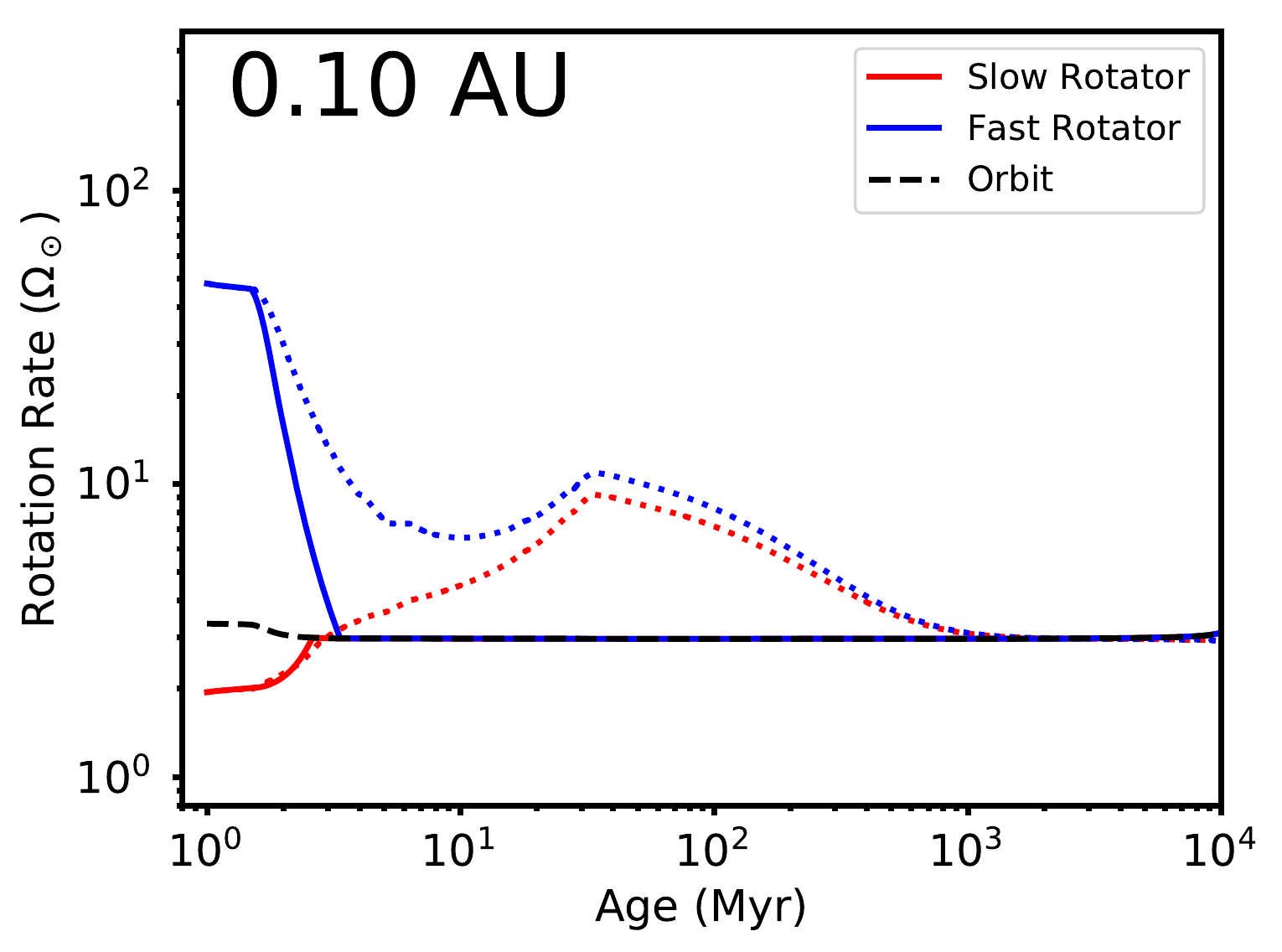}
\includegraphics[trim = 0mm 0mm 0mm 0mm, clip=true,width=0.42\textwidth]{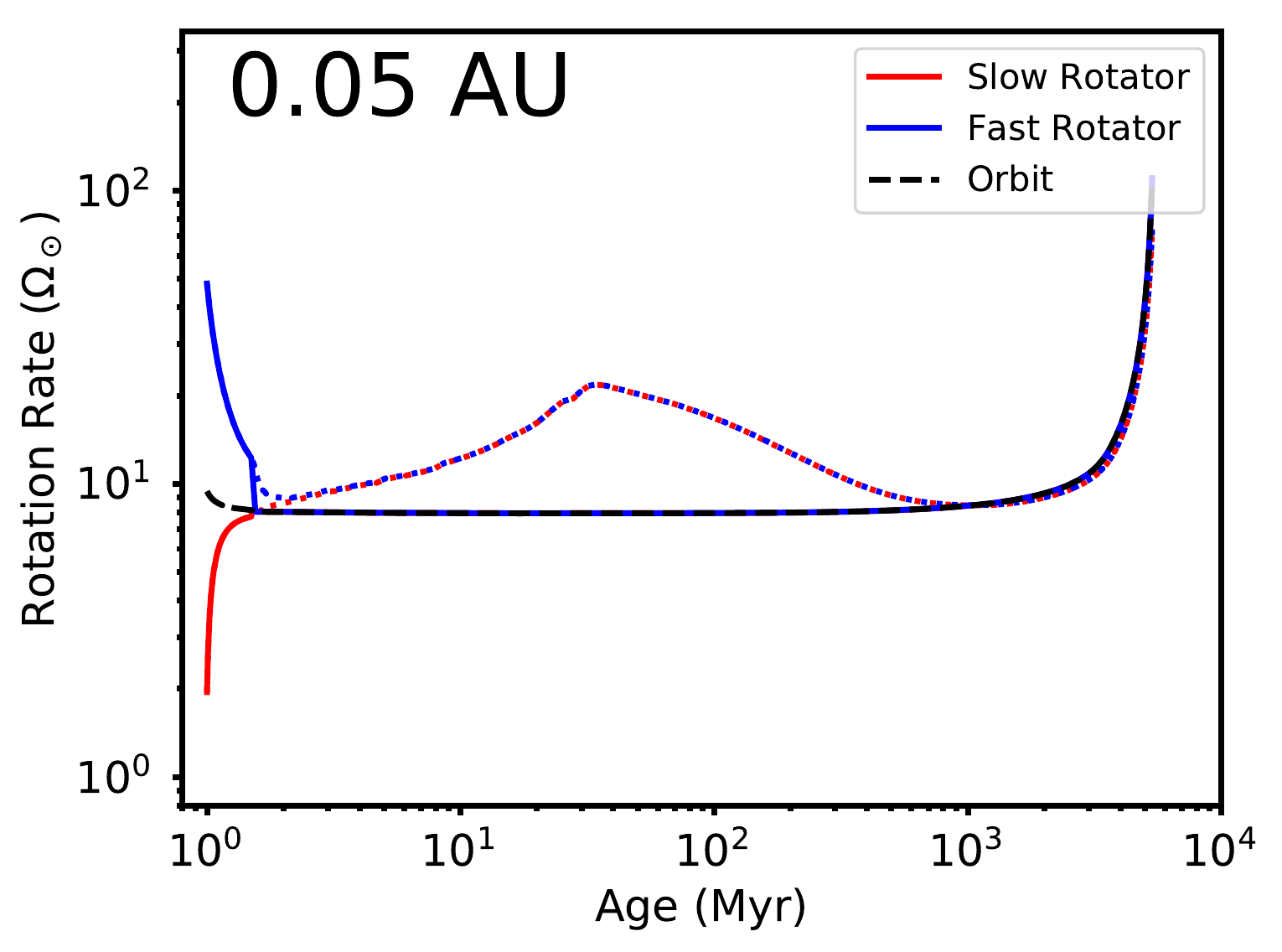}
\includegraphics[trim = 0mm 0mm 0mm 0mm, clip=true,width=0.42\textwidth]{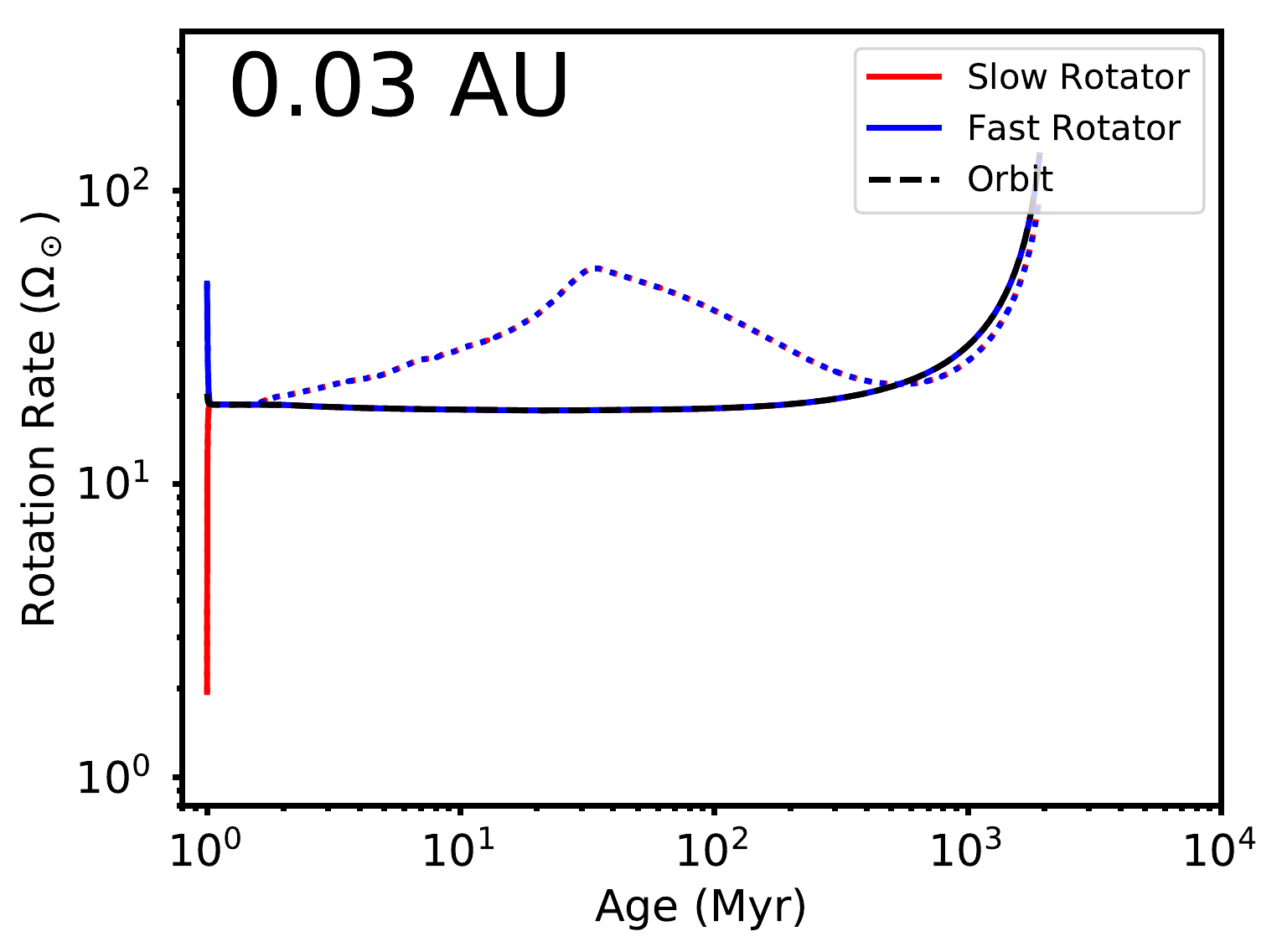}
\caption{
Rotational evolution of two solar mass stars in tight binary systems with different initial orbital separations.
In each system, the two stars start with very different initial rotation rates.
The initial orbital separations are written in the top left of each panel.
The dashed black lines show the orbital angular velocity.
In several panels, the slow rotator line is mostly covered by the fast rotator line.
}
 \label{fig:binaryrotationtracks}
\end{figure*}

We solve this system of ODEs using the implicit multi-step Rosenbrock solver described in \citet{sandu1997benchmarking} and Appendix~H of \citet{Johnstone18}, where their $\mathbf{n}$ is replaced by \mbox{$\mathbf{X} = [ \Omega_\mathrm{env,1} , \Omega_\mathrm{core,1} , \Omega_\mathrm{env,2} , \Omega_\mathrm{core,2} , a_\mathrm{orb} ]^\mathrm{T}$}.
In this case, we calculate the Jacobian matrix, \mbox{$\mathbf{J} = d \mathbf{F} / d \mathbf{X}$}, where \mbox{$\mathbf{F} = d \mathbf{X} / dt$}, numerically by perturbing each variable in $\mathbf{X}$ by small amounts.
The lengths of each timestep are calculated automatically by the solver.
A difficulty arises however when the orbital separations are small: due to how rapidly angular momentum is exchanged when dynamical tides dominate, small orbital separations lead to very large tidal synchronisation torques and therefore very short synchronisation timescales.
This causes our numerical solver to require very short timesteps to maintain accuracy and stability, which become restrictively short for orbital separations smaller than approximately 0.1~AU.
We avoid this problem by assuming that the stars are perfectly synchronised when the synchronisation timescale, given by \mbox{$I_\star |\Omega_\mathrm{env} - \Omega_\mathrm{orb}|/\tau_\mathrm{ts}$}, is less than 0.01~Myr for both stars and when the values of $\Omega_\mathrm{env}$ of both stars are within 0.1\% of $\Omega_\mathrm{orb}$.
When these conditions are met, the stars are anyway in almost perfect tidal synchronisation, so the assumption is reasonable. 
As described in Appendix~\ref{appendix:modifiedmodel}, this requires small modifications to the equations that are solved. 
By making this assumption, the tidal synchronisation torque is not needed in the timestep update and the restrictions on the timestep size are made significantly less severe. 
The above conditions are tested at the start of each timestep.

To test our model, we calculate three rotational evolution models between 1~Myr and 10~Gyr for solar mass stars with different initial rotation rates.
For the initial rotation rates at 1~Myr, we choose the values used by \mbox{\citet{Tu15}} for their slow, medium, and fast rotator models.
The initial rotation rates in these models are 1.9, 6.5, and 48 $\Omega_\odot$.
These tracks represent stars at the 10th, 50th, and 90th percentiles of the rotational distributions for single stars.
The rotation tracks are shown in Fig.~\ref{fig:solarrotation}.
For all tracks, the solid and dotted lines show the envelope and core rotation tracks, and the black dashed lines show the models from \mbox{\citet{Tu15}} for comparison.
Clearly, our models match well the comparison models.
In the remainder of this paper, when we refer to `slow', `medium', and `fast' rotators, we mean stars that have initial rotation rates equal to those used in these single star models.

\subsection{Results}

\begin{figure}
\centering
\includegraphics[trim = 0mm 0mm 0mm 0mm, clip=true,width=0.47\textwidth]{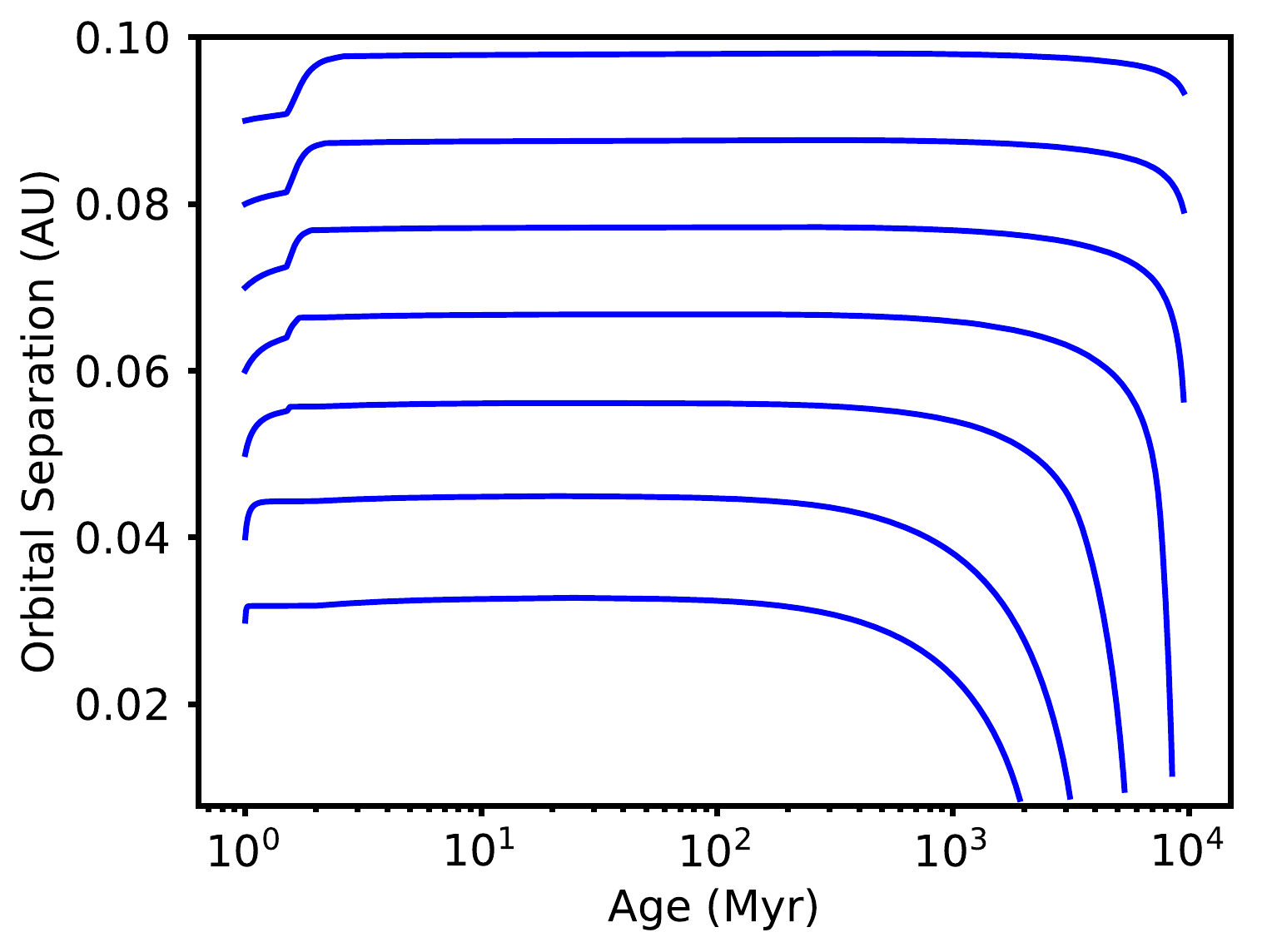}
\includegraphics[trim = 0mm 0mm 0mm 0mm, clip=true,width=0.49\textwidth]{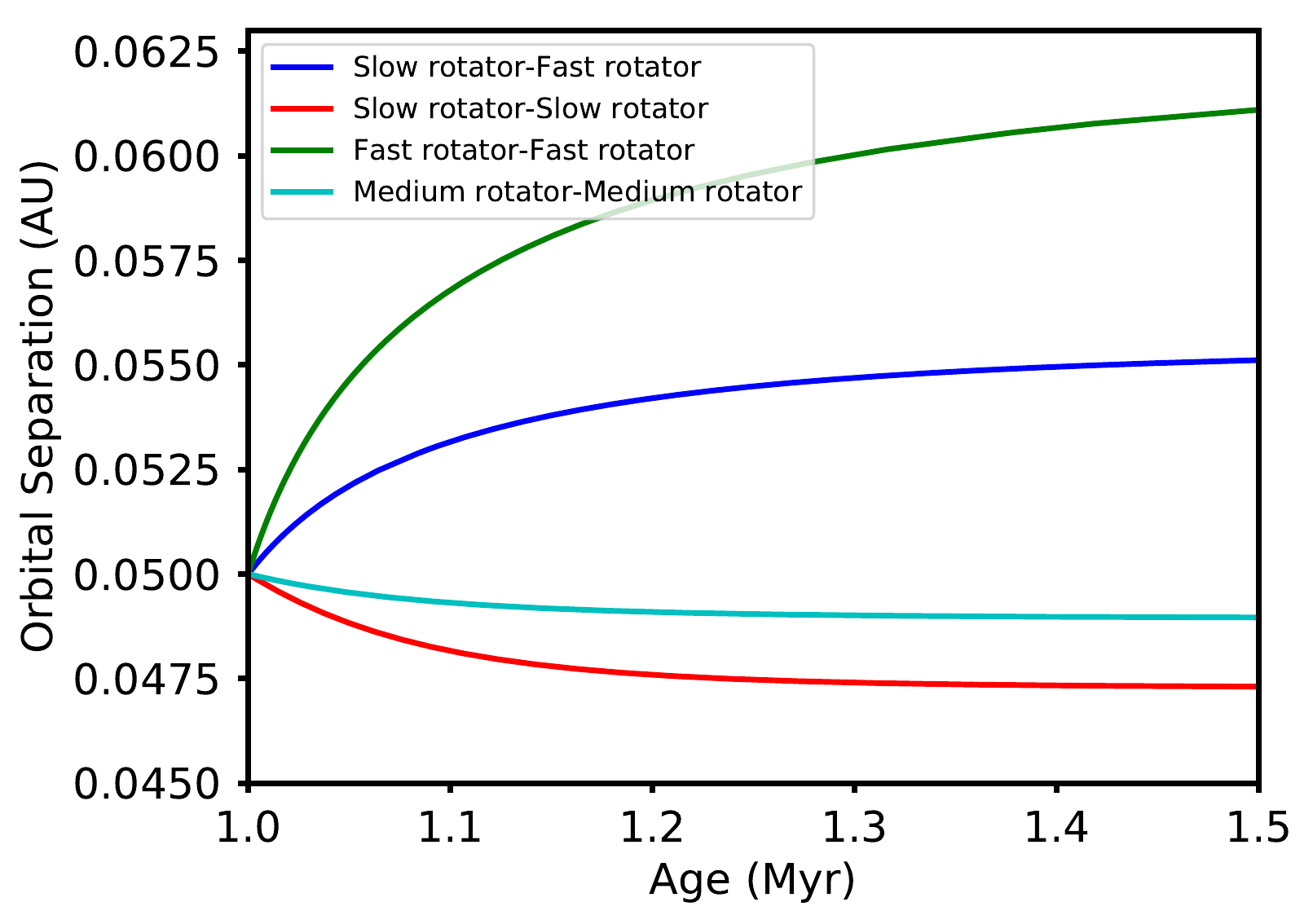}
\caption{
Evolution of the orbital separation of tight binary systems composed of two solar mass stars.
The upper panel shows the evolutionary tracks for the systems shown in Fig.~\ref{fig:binaryrotationtracks}, which consist of one initially fast rotator and one initially slow rotator.
The systems differ only in their initial orbital separations.
The lower panel shows the orbital evolution in the first 0.5~Myr for four systems with starting initial separations of 0.05~AU.
These systems differ in the initial rotation rates of their two stars, as indicated in the legend. 
}
 \label{fig:separationtracks}
\end{figure}

\begin{figure}
\centering
\includegraphics[trim = 0mm 0mm 0mm 0mm, clip=true,width=0.49\textwidth]{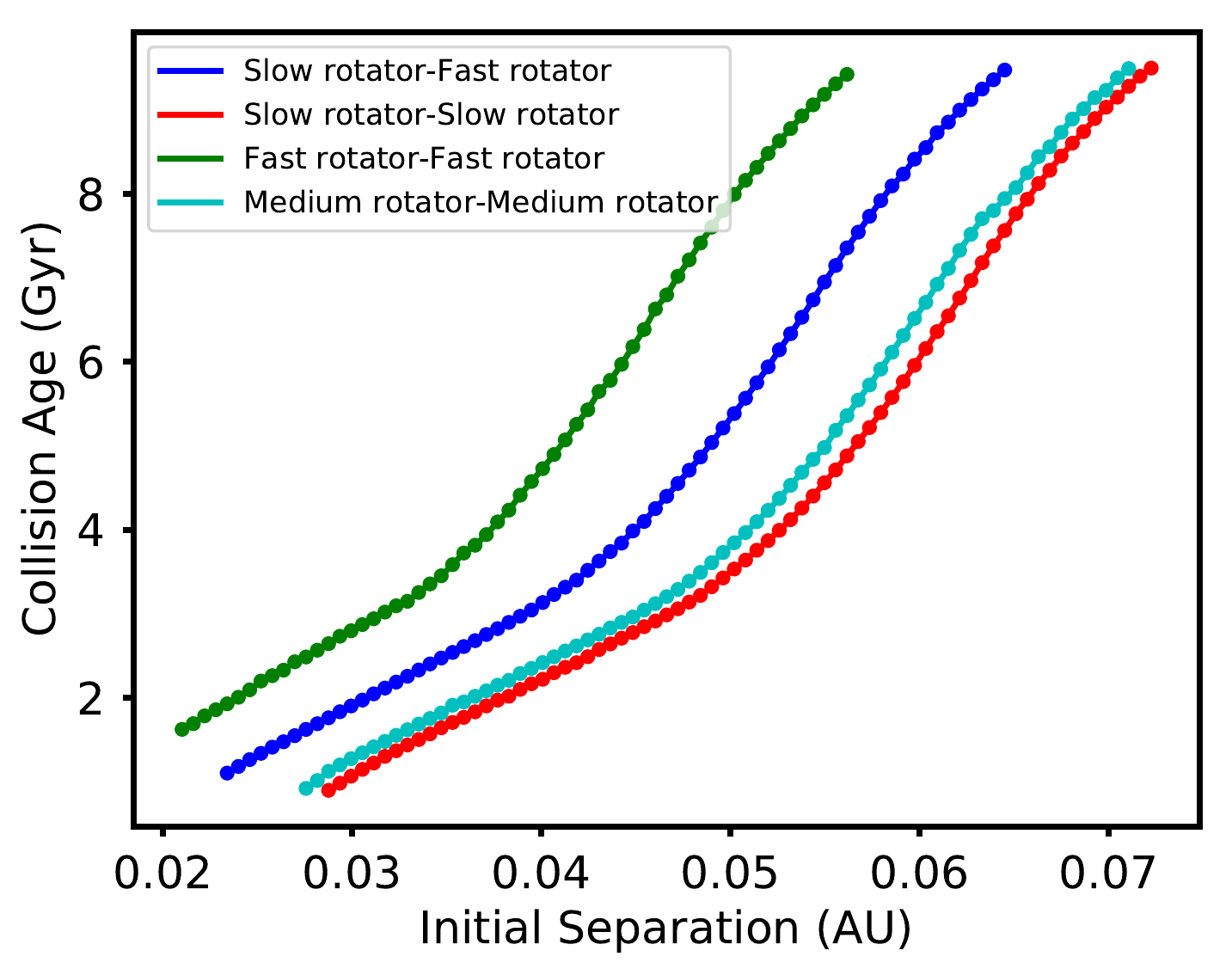}
\caption{
Age at which the two stars in tight binary systems collide as a function of their initial orbital separations.
This takes place because their decay due to the removal of angular momentum from the binary system by stellar winds.
The four cases shown differ only in the initial rotation rates of the two stars, as indicated in the legend.
The points show the results from individual roational evolution models.
}
 \label{fig:collisionage}
\end{figure}

\begin{figure*}
\centering
\includegraphics[trim = 0mm 0mm 0mm 0mm, clip=true,width=0.41\textwidth]{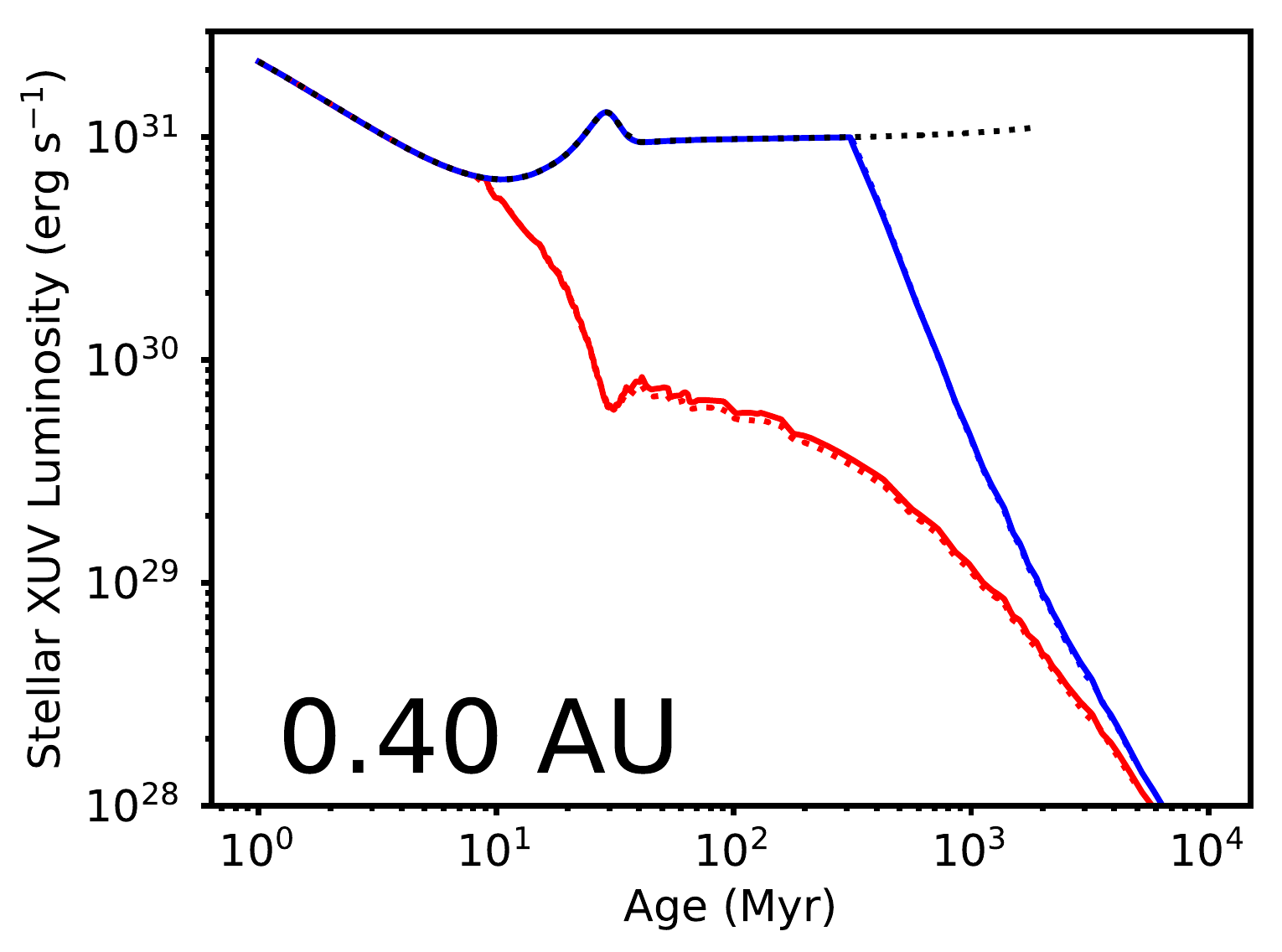}
\includegraphics[trim = 0mm 0mm 0mm 0mm, clip=true,width=0.41\textwidth]{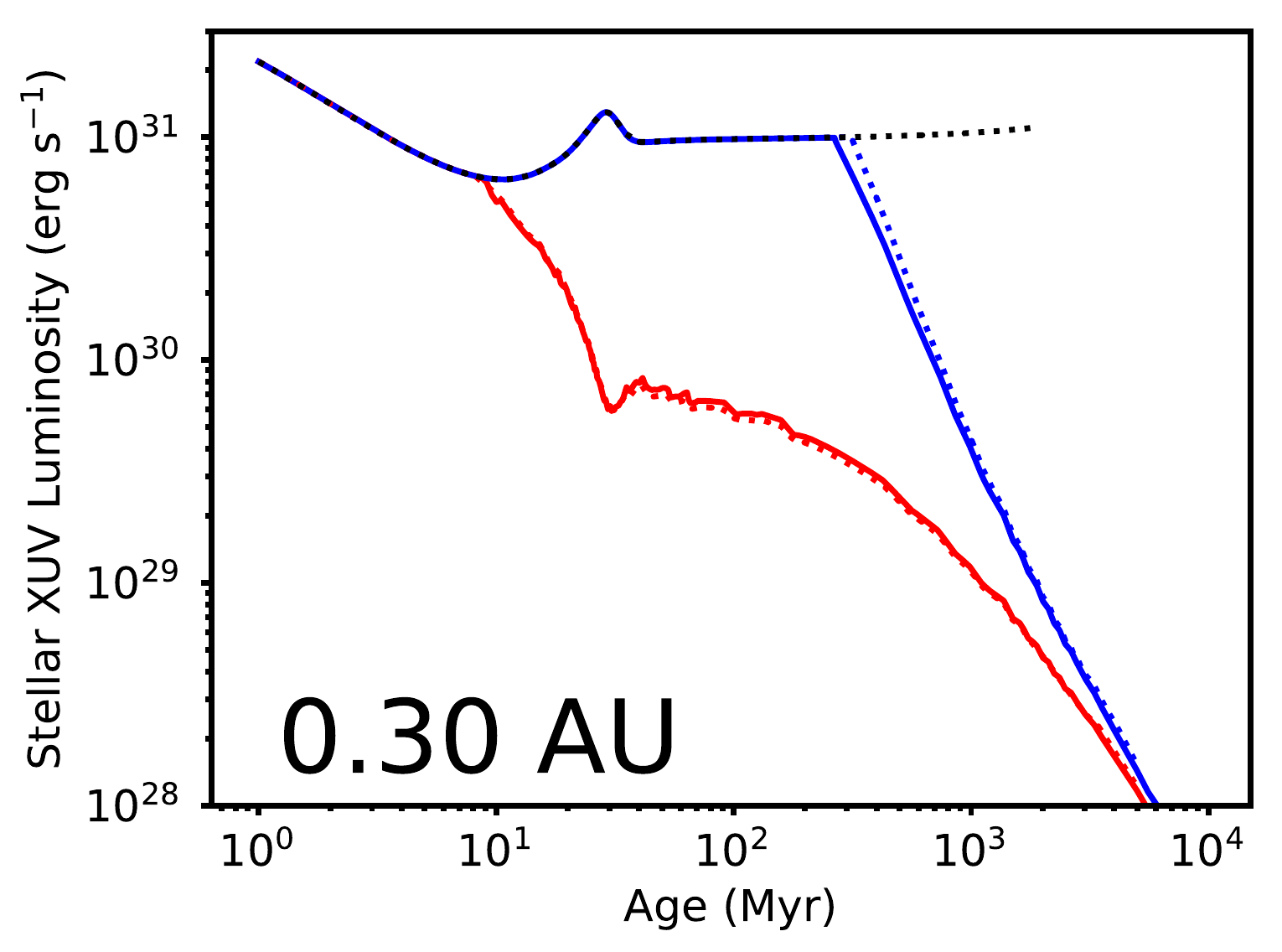}
\includegraphics[trim = 0mm 0mm 0mm 0mm, clip=true,width=0.41\textwidth]{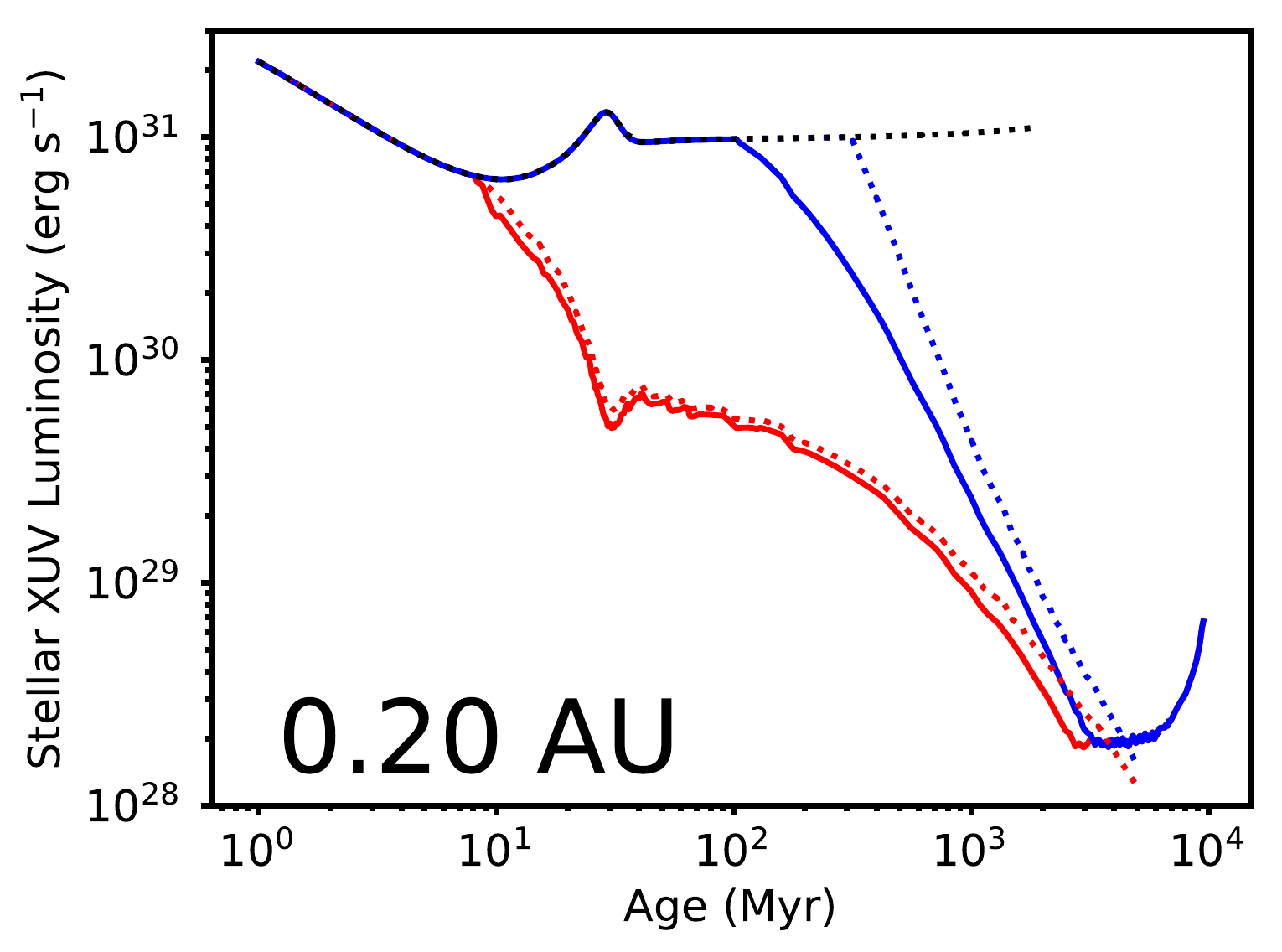}
\includegraphics[trim = 0mm 0mm 0mm 0mm, clip=true,width=0.41\textwidth]{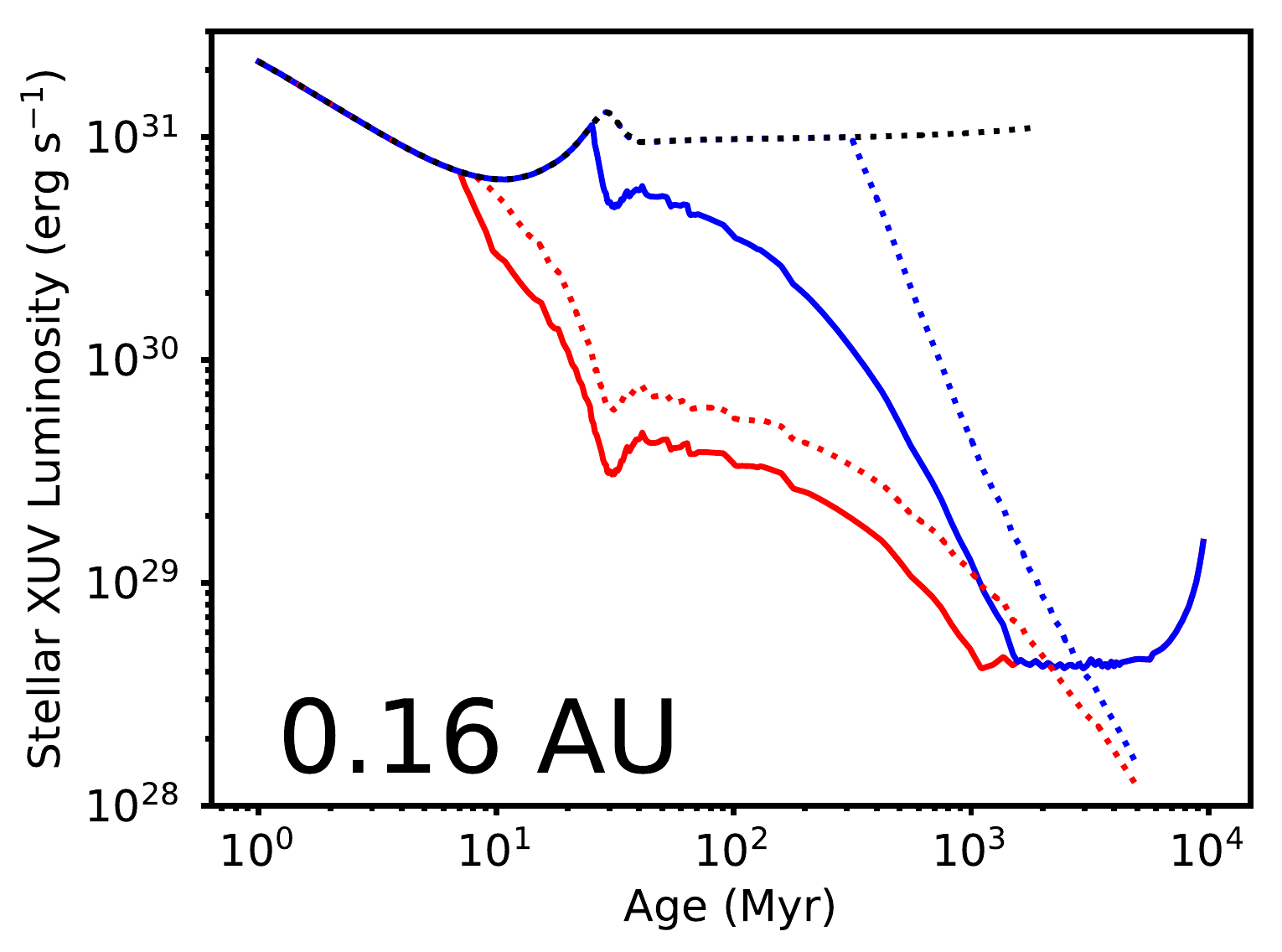}
\includegraphics[trim = 0mm 0mm 0mm 0mm, clip=true,width=0.41\textwidth]{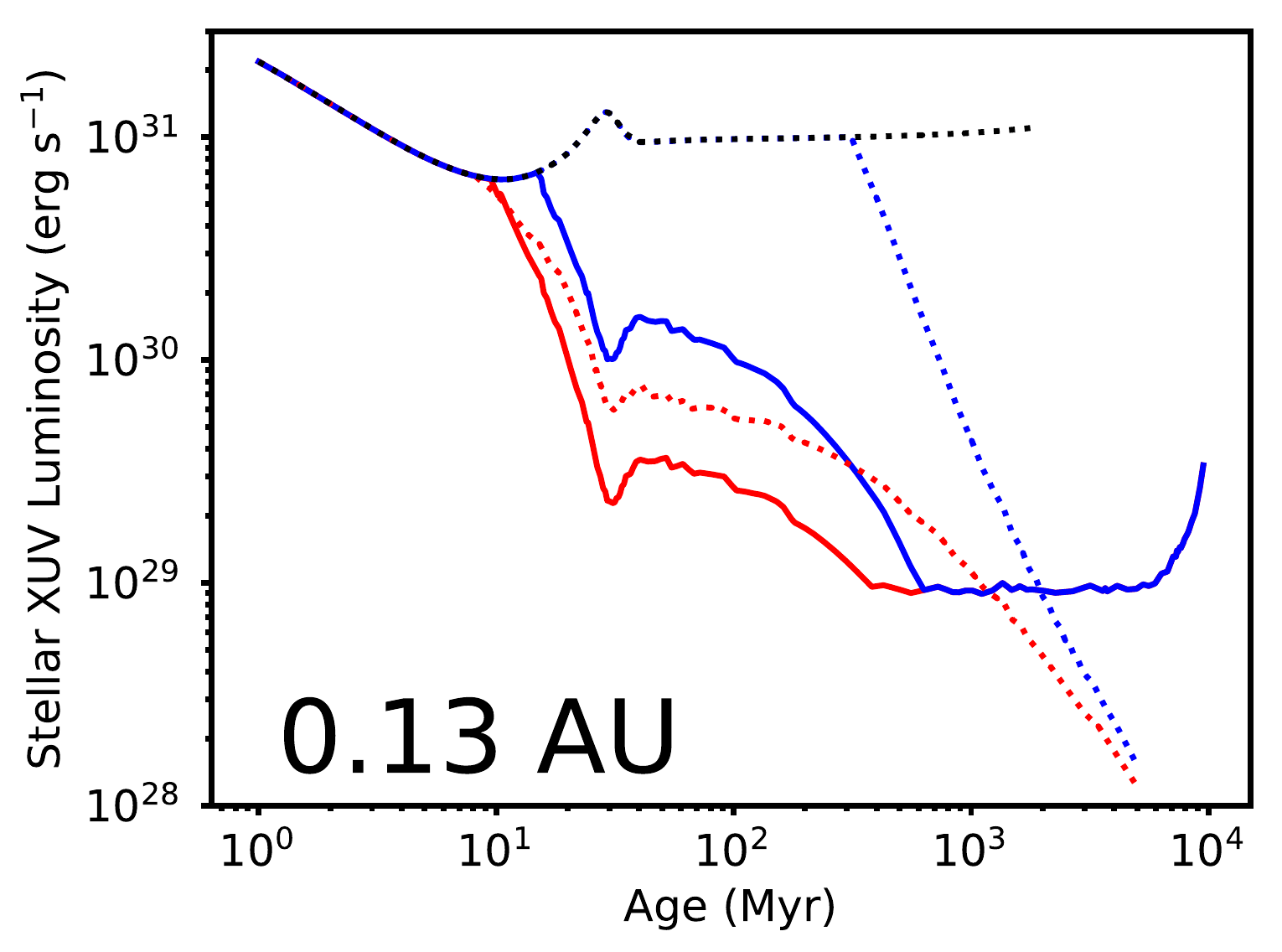}
\includegraphics[trim = 0mm 0mm 0mm 0mm, clip=true,width=0.41\textwidth]{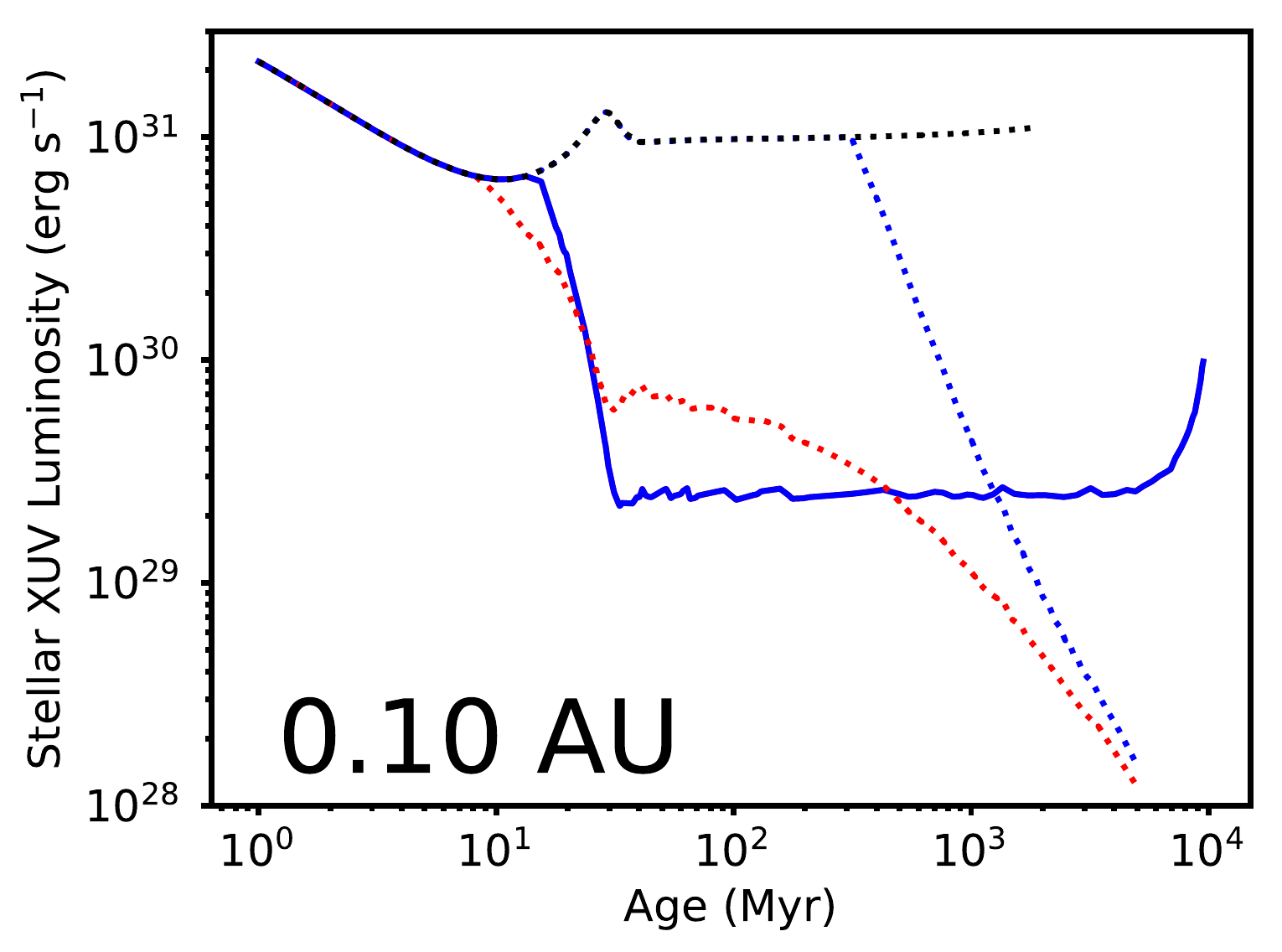}
\includegraphics[trim = 0mm 0mm 0mm 0mm, clip=true,width=0.41\textwidth]{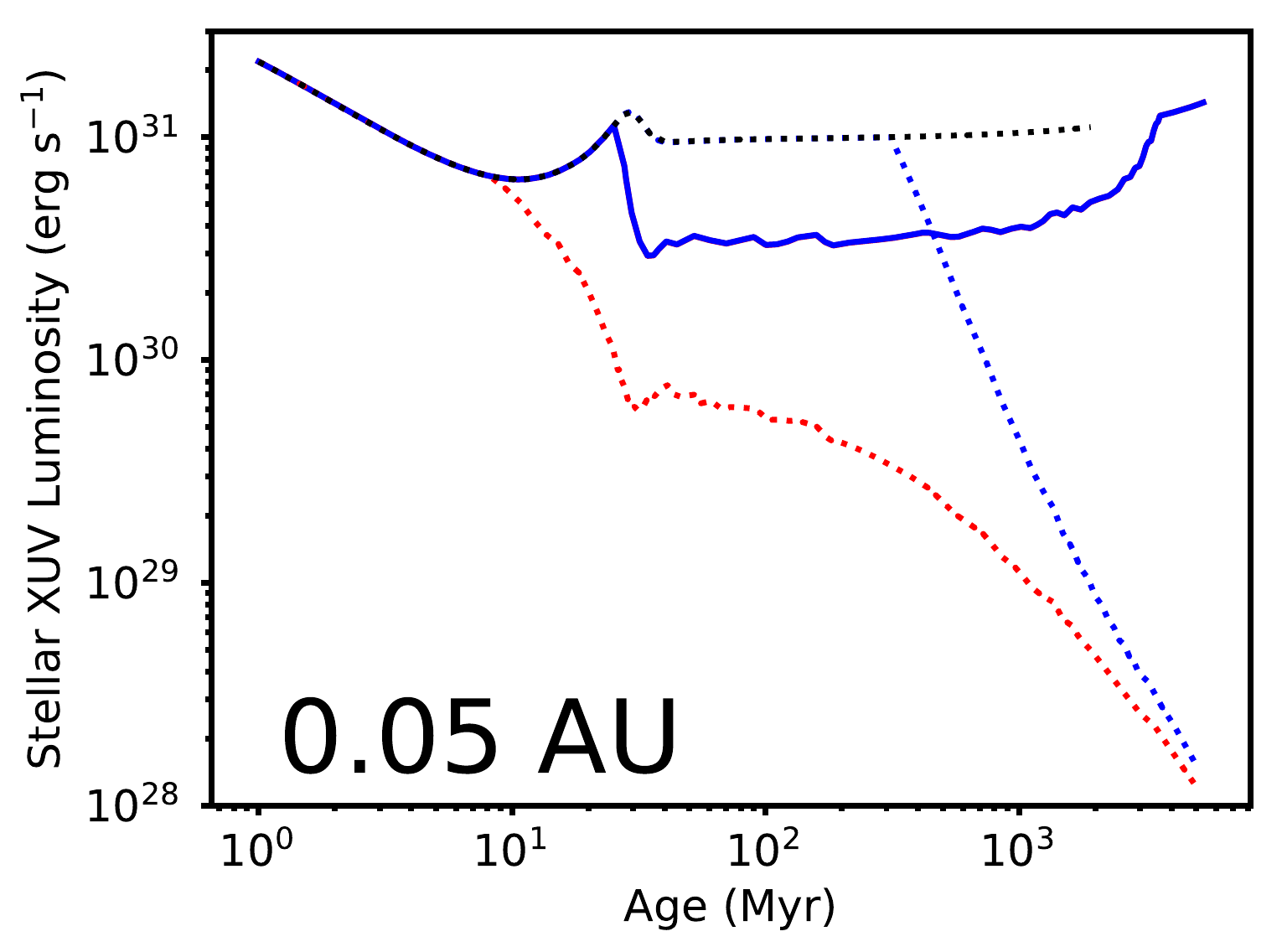}
\includegraphics[trim = 0mm 0mm 0mm 0mm, clip=true,width=0.41\textwidth]{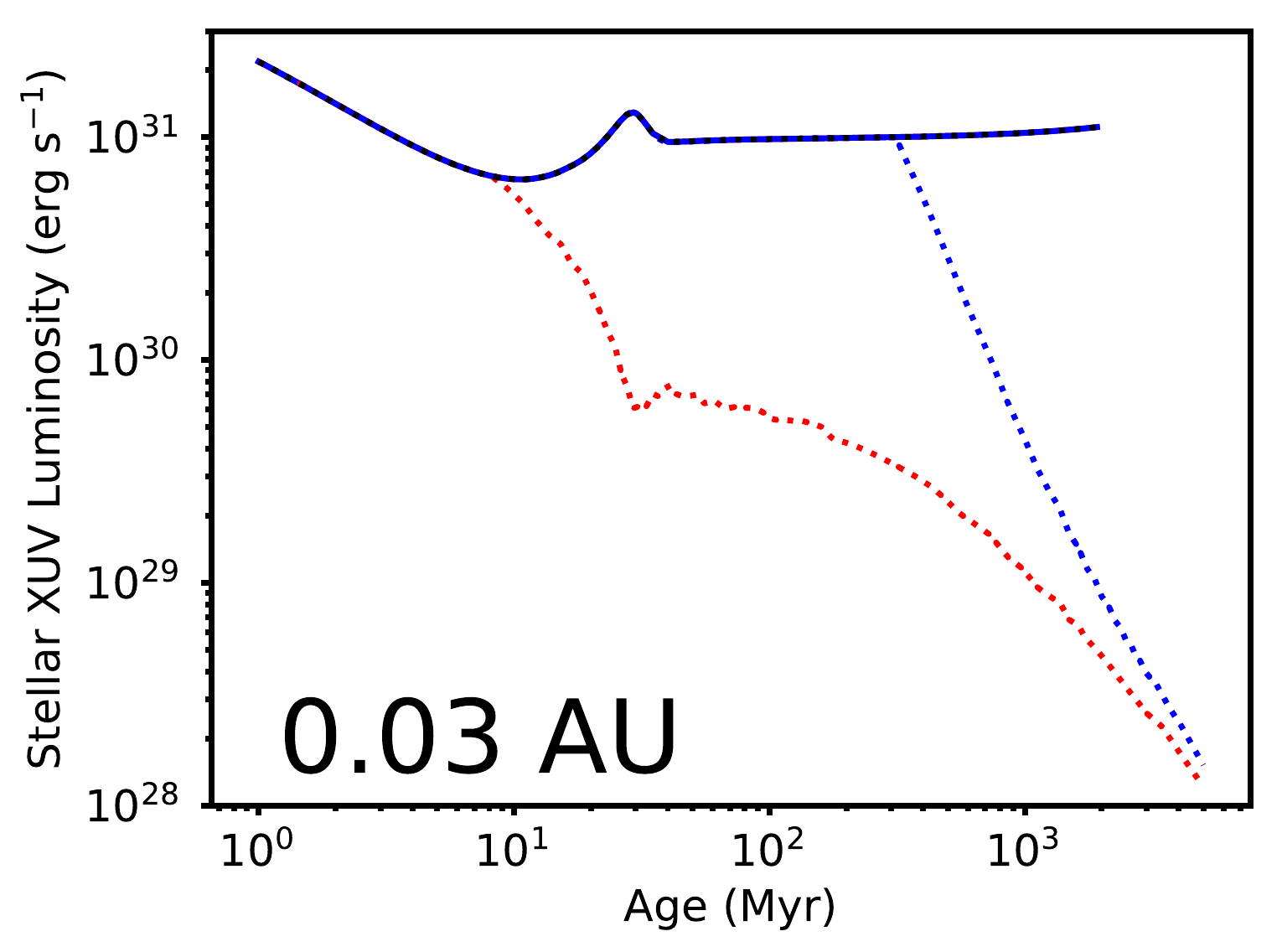}
\caption{
Radiation evolution of two solar mass stars in tight binary systems with different initial orbital separations.
In each system, the two stars start with very different initial rotation rates, with red and blue indicating the tracks for slow and fast rotators respectively.
The initial orbital separations are written in the botom left of each panel.
The evolutionary tracks end when the two stars merge due to the removal of angular momentum in the stellar wind.
The evolutionary tracks that these stars would follow in single systems are shown as dotted lines.
The black dotted line shows the saturation XUV luminosity.
}
 \label{fig:binaryradiationtracks}
\end{figure*}

\begin{figure*}
\centering
\includegraphics[trim = 0mm 0mm 0mm 0mm, clip=true,width=0.49\textwidth]{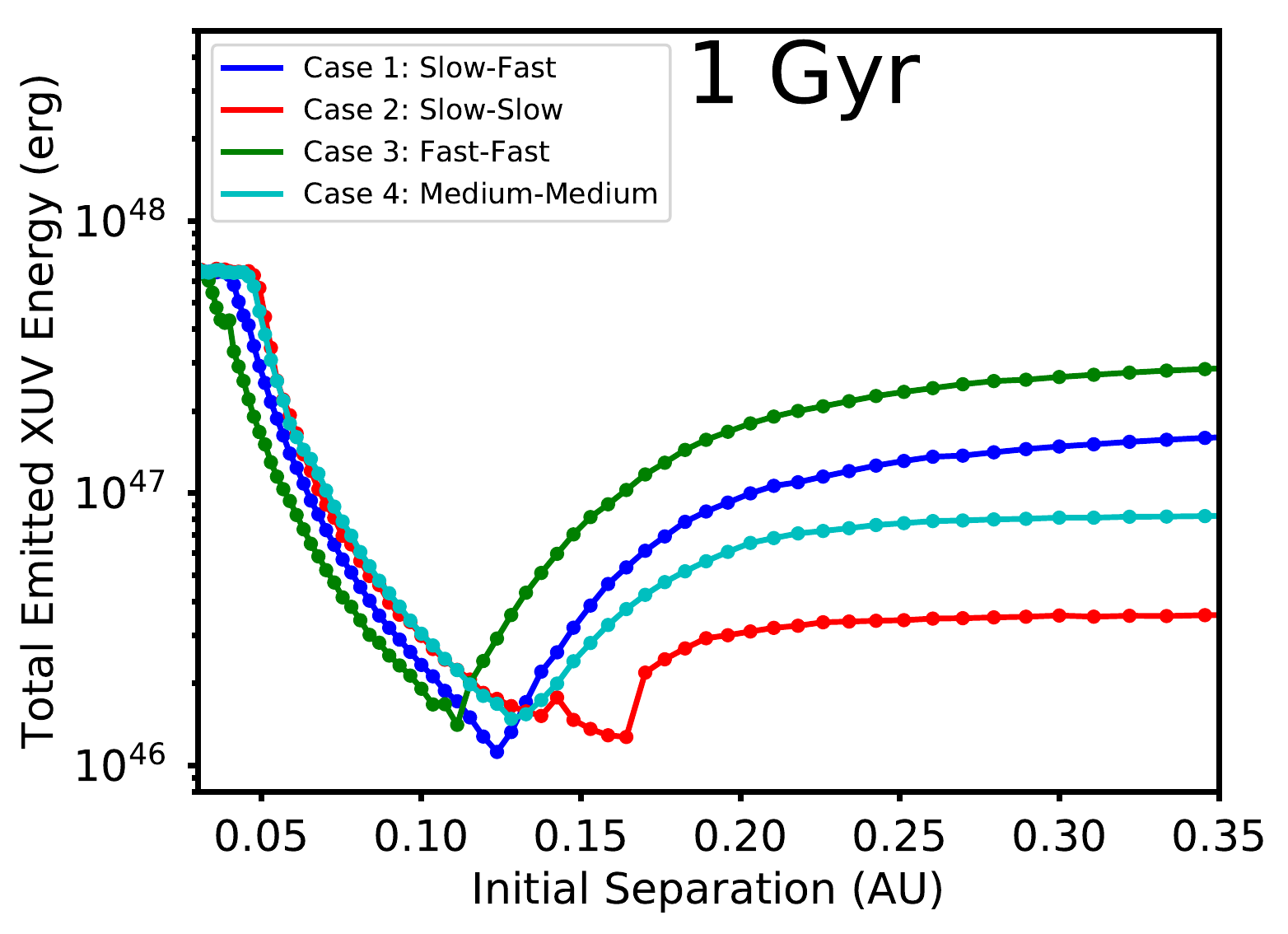}
\includegraphics[trim = 0mm 0mm 0mm 0mm, clip=true,width=0.49\textwidth]{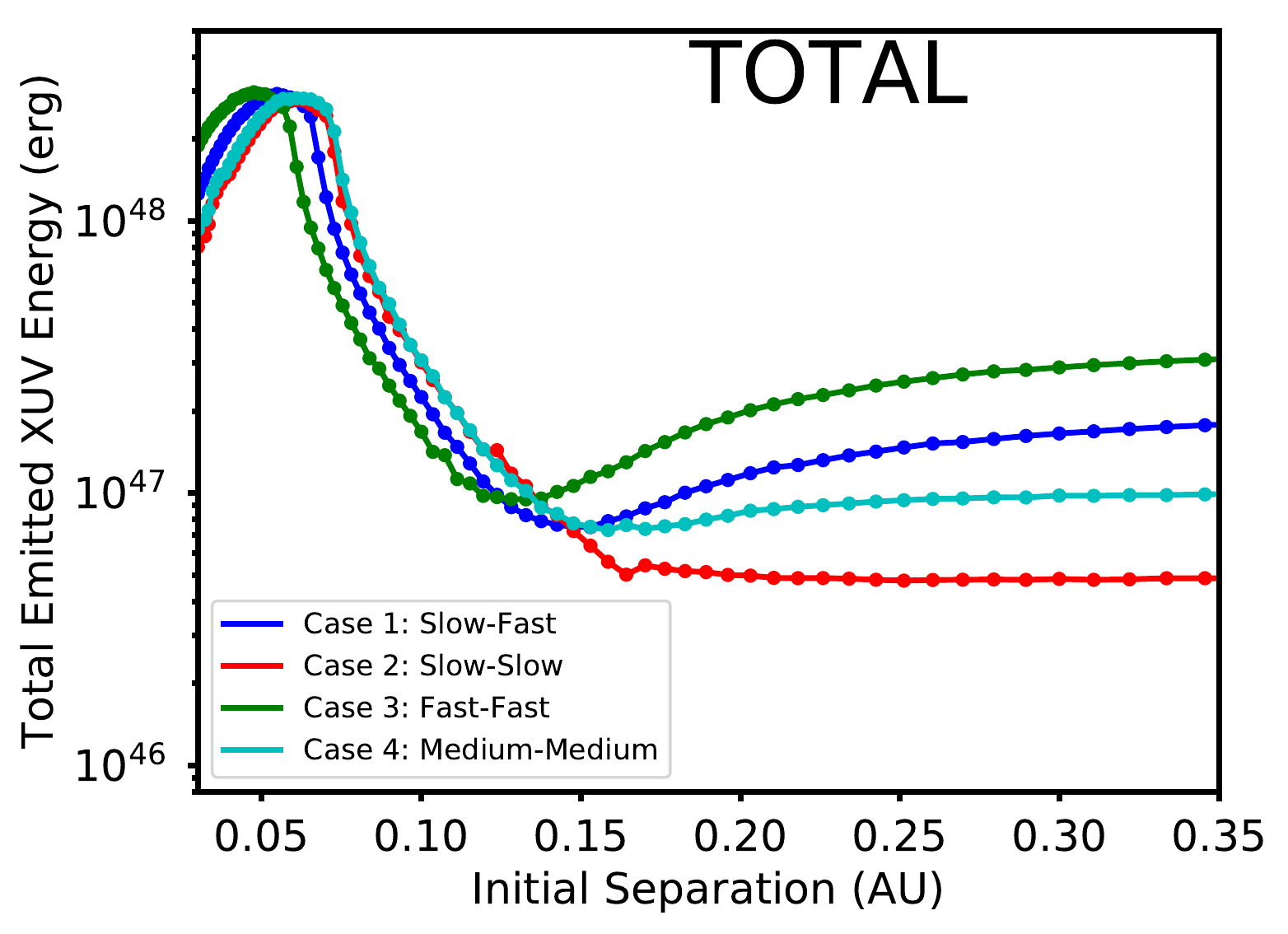}
\includegraphics[trim = 0mm 0mm 0mm 0mm, clip=true,width=0.49\textwidth]{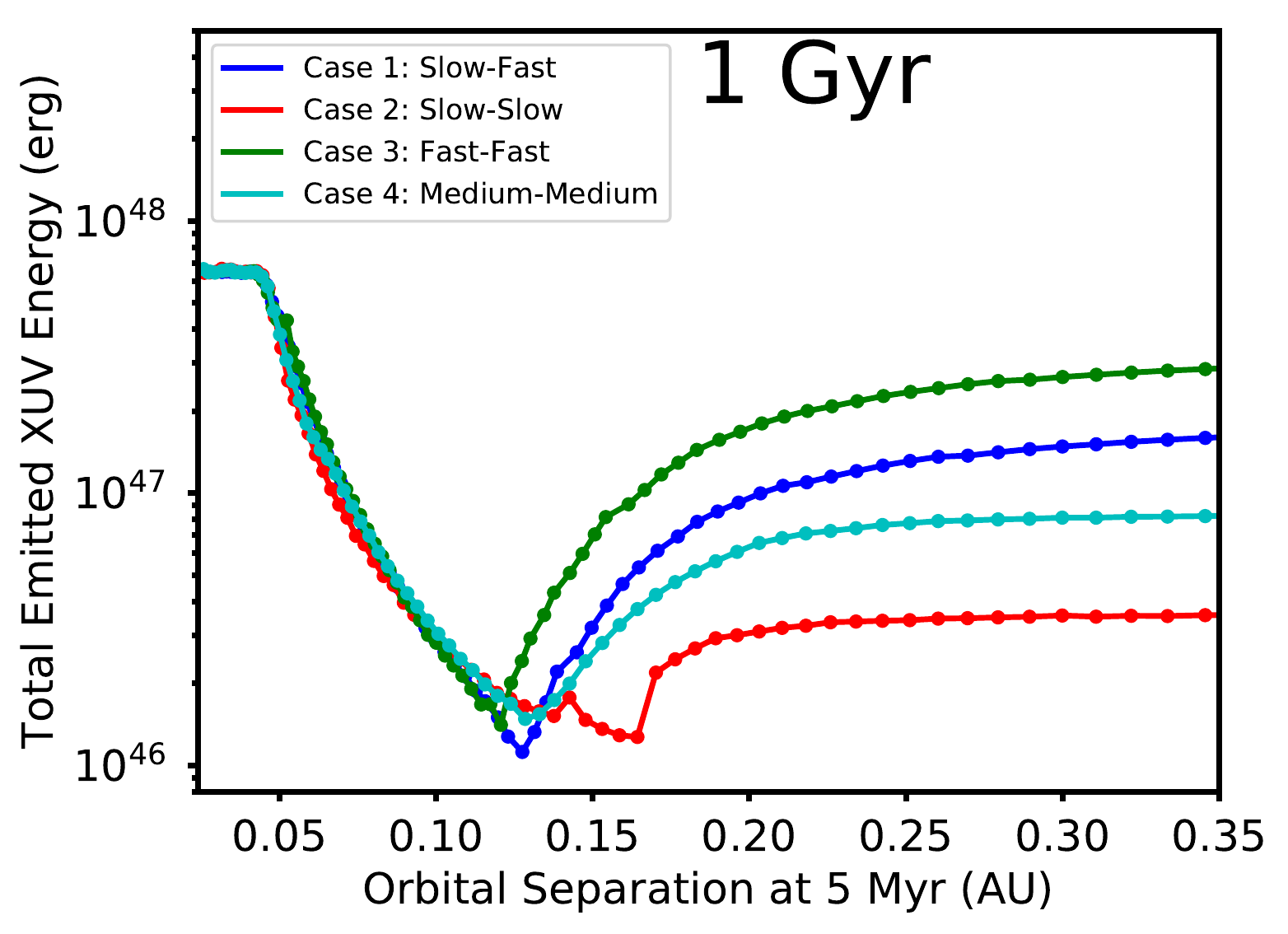}
\includegraphics[trim = 0mm 0mm 0mm 0mm, clip=true,width=0.49\textwidth]{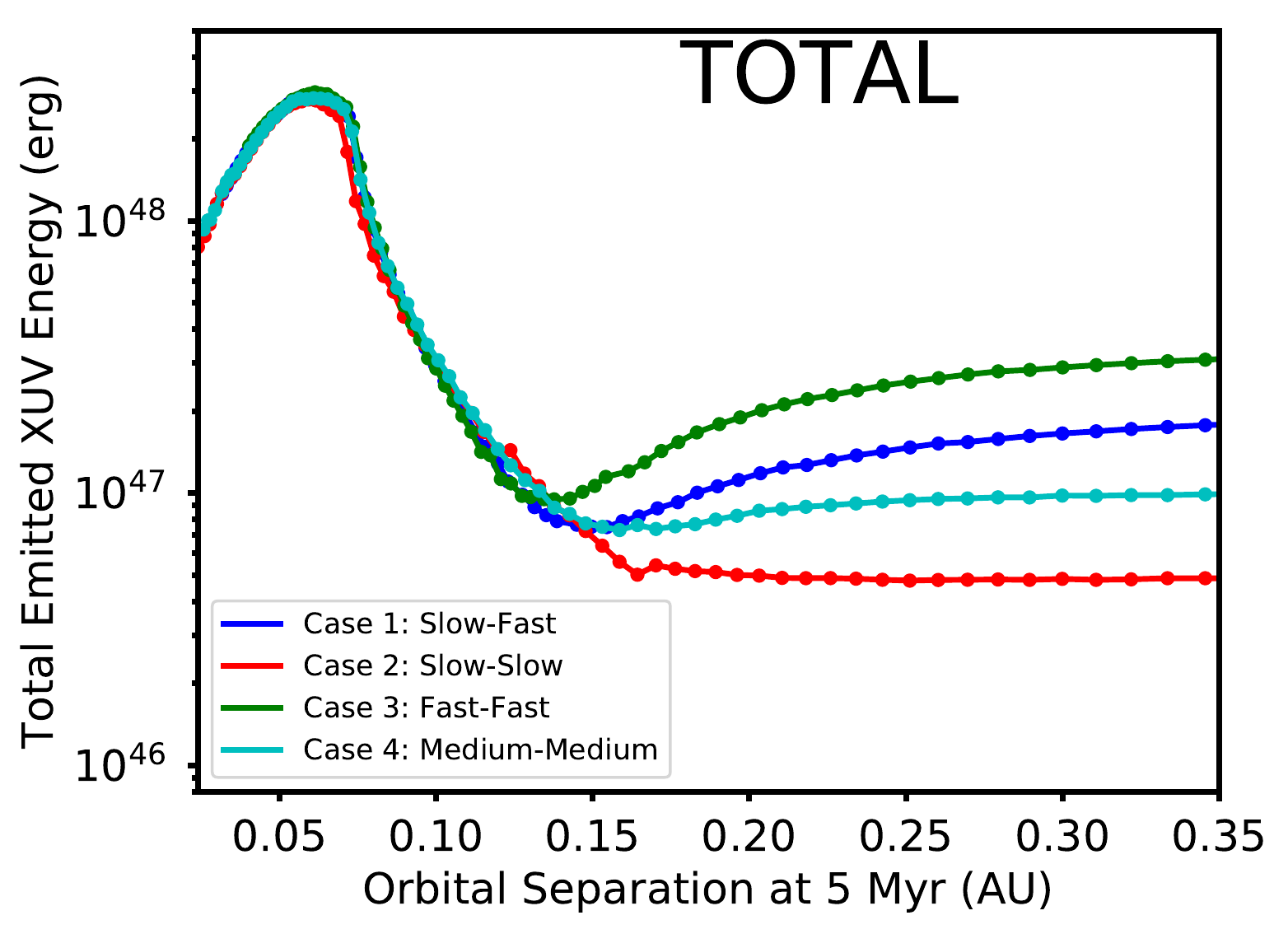}
\caption{
Total emitted XUV energy by both stars between ages of 1~Myr and either 1~Gyr (left~column) or the end of their lifetimes (right~column).
The upper panels show the values as a function of the initial orbital separation, and the lower panels show the values as a function of the orbital separation at 5~Myr.
The four cases shown in each panel differ only in the initial rotation rates of the two stars, as indicated in the legend.
}
 \label{fig:binarytotalradiation}
\end{figure*}

\begin{figure}
\centering
\includegraphics[trim = 0mm 0mm 0mm 0mm, clip=true,width=0.49\textwidth]{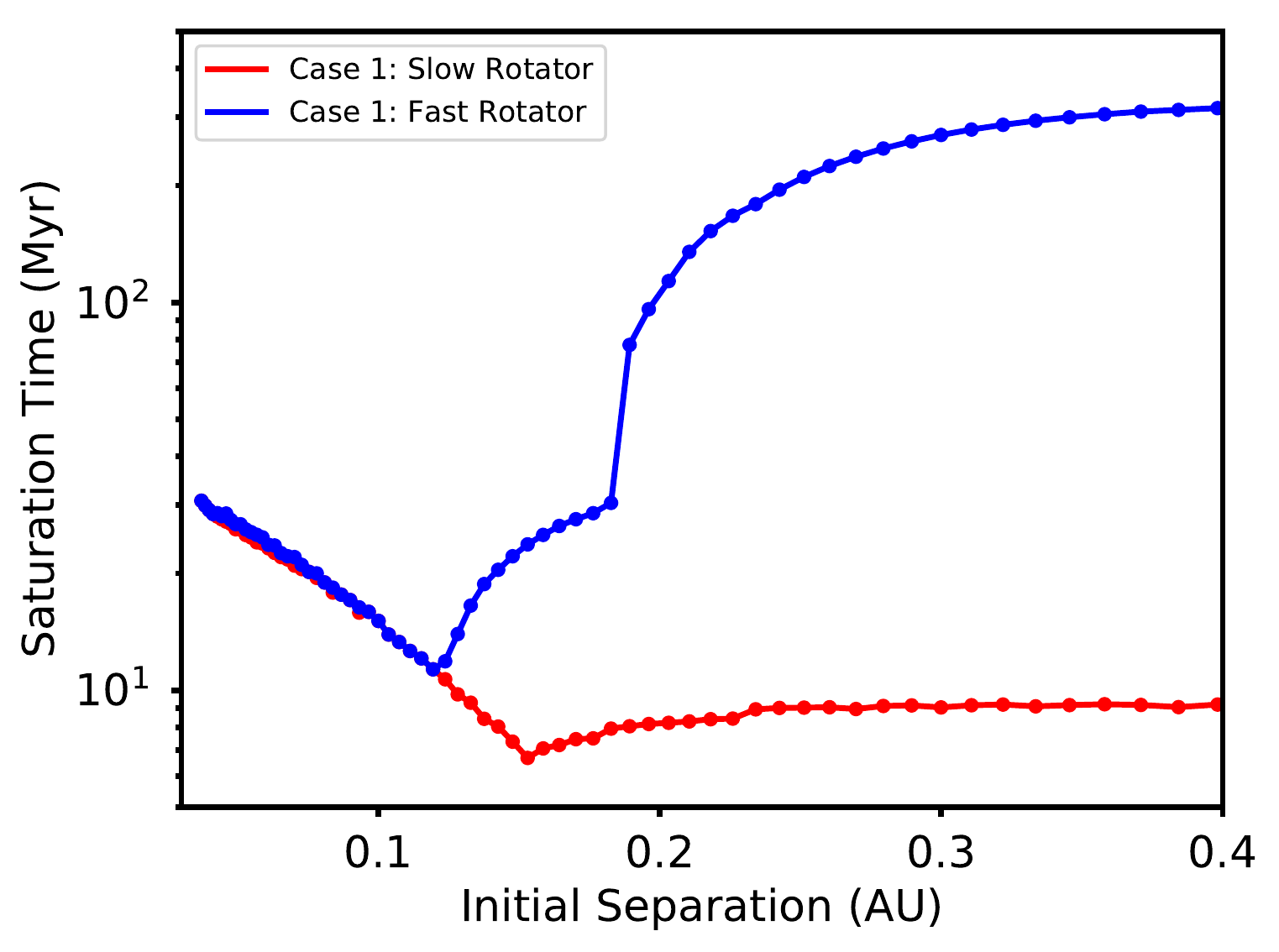}
\includegraphics[trim = 0mm 0mm 0mm 0mm, clip=true,width=0.49\textwidth]{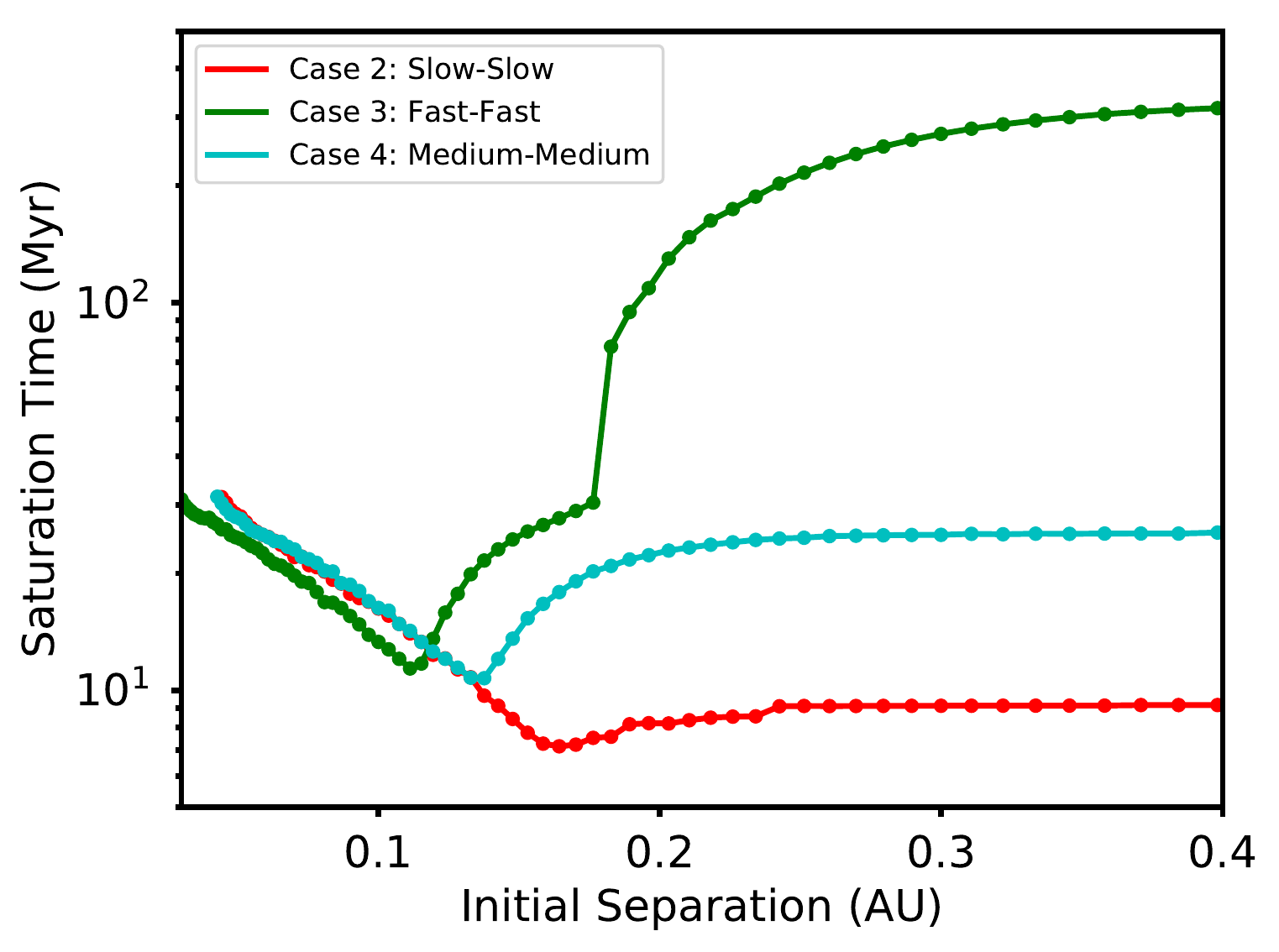}
\caption{
Ages at which stars drop out of saturation as a function of initial orbital separation for all four cases.
In the upper panel, we show this dependence for both stars in Case~1, which has one initially slow rotator and one initially fast rotator.
In the lower panel, we show this dependence for the other three cases.
}
 \label{fig:binarysaturationage}
\end{figure}

Since we only consider solar mass stars in this paper, there are three input parameters in our model: these are the initial rotation rates of the two stars, and the initial orbital separation.
We consider four cases, each one differing in the initial rotation rates of the two stars.
In Case~1, one star starts as a slow rotator and the other as a fast rotator.
In Case~2, both stars start as slow rotators.
In Case~3, both stars start as fast rotators.
In Case~4, both stars start as medium rotators.
For each case, we calculate a series of evolutionary models with different initial orbital separations.

In Fig.~\ref{fig:binaryrotationtracks}, we show the rotational evolution tracks for Case~1, with different initial orbital separations.
In the models with an initial orbital separations greater than $\sim$0.3~AU, the tidal synchronisation torque has only a small influence on the rotational evolution and the rotation rates evolve almost as they would in single systems.
At orbital separations greater than $\sim$0.1~AU, tidal torques require a significant amount of time ($\sim$500~Myr or longer) to bring the stars into perfect synchronisation, and the rotation tracks are often quite complex with different processes dominating at different times. 
For example, in the 0.13~AU case the slow rotator becomes synchronised with the orbit at an age of approximately 2~Myr, but then due to the star's contraction the tidal torques become weaker and at an age of 20~Myr the star stops being perfectly synchronised and spins up, before spinning down and becoming tidally locked again at approximately 350~Myr.
At initial separations less than $\sim$0.1~AU, tidal effects dominate entirely and all stars become tidally locked almost immediately.

In the models with small initial orbital separations, the tidal synchronisation torques have an important effect on the orbital separations.
This is due to the stellar winds removing angular momentum from the system; in order to compensate for this angular momentum removal and keep the rotation rates of the two stars synchronised, the tidal synchronisation torques remove angular momentum from the orbit, causing the orbital separtion to decay. 
The rapid increases in the rotation rates at the end of the evolution tracks in these cases is due to the decaying of the orbit and the corresponding increase in the orbital angular velocity.
The evolution of the orbital separation for several of these models is shown in the upper panel of Fig.~\ref{fig:separationtracks}.
In the cases with the smallest initial orbital separations, the winds remove so much angular momentum that the two stars eventually collide. 
These stars eventually merge to form a single star with a mass of \mbox{$M_1+M_2$}.

It should be expected that tidally locked stars rotate slightly slower than the orbital rotation rate since when \mbox{$\Omega_\mathrm{env}=\Omega_\mathrm{orb}$} the tidal torque vanishes, and therefore stars should spin down due to stellar wind torques. 
Tidally locked stars should rotate with the rate at which the tidal torque balances the wind torque.
This sub-synchronous rotation can be seen in the models of \citet{Zuluaga16} and would be seen in our models if we had not included dynamical tides. 
However, given how strong tidal synchronisation torques become when dynamical tides are considered, this effect is negligible and tidally locked stars rotate at almost exactly the orbital rotation rate.
Interestingly, \citet{Lurie17} found from Kepler observations of tight binaries (see their Fig.~6) some systems in which the measured rotation periods were $\sim$13\% slower than the orbital periods, which they suggest is due to differential rotation and high latitude starspots.
Alternatively, this could suggest that our tidal sychronisation torques are too strong, and in reality the sub-synchronous rotation rates seen by \citet{Lurie17} are simply due to the balance between tidal torques and wind torques being where $\Omega_\mathrm{env}$ is slightly below $\Omega_\mathrm{orb}$.
In support of the stronger torques however are the measurements of \citet{2010ApJ...723..285H} who used observations of exoplanet systems and of binary star systems to measure tidal dissipation factors ($\sigma_\star$ in Eqn.~\ref{eqn:disstimeEQ}) assuming equilibrium tides.
They found values of $\sigma_\star$ for exoplanet systems that are more than two orders of magnitude smaller than the value derived from binary star systems, which they suggest is due to the binary systems considered being close to synchronisation and therefore affected by strong dynamical tides (see the end of their Section~5).

An interesting feature of the orbital evolution shown in Fig.~\ref{fig:separationtracks} is that at the beginnings of some of the tracks, the orbital separations in fact increase.
This is shown in more detail in the lower panel of Fig.~\ref{fig:separationtracks} for the four cases, each with initial orbital separations of 0.05~AU.
The four models differ in the initial rotation rates of the two stars.
In the two extreme cases, the stars start out both as slow rotators (red line) and as fast rotators (green line). 
When both stars are slow rotators, the orbital separation quickly decreases as the tidal synchronisation torques transfer angular momentum from the orbit to the rotations of the two stars in order to spin them up.
When both stars are fast rotators, the orbital separation instead increases as the tidal synchronisation torques transfer angular momentum in the opposite direction to spin the stars down.
This initial increase in the orbital period was recently used by \citet{Fleming18} to explain why so few circumbinary planets have been observed orbiting close binary star systems.
Observationally, it appears that circumbinary planets tent to orbit just exterior to the dynamical stability limit, inside of which the orbits of planets are dynamically unstable (\citealt{WinnFabrycky15}), and this can be understood as a result of planetary inward migration in the circumstellar gas disk stopping at the dynamical stability limit where the disk is truncated.
\citet{Fleming18} suggest that as the orbital separation of the two stars increases due to the processes described above, the dynamical stability limit moves outwards, and the orbits of planets that were previously stable become unstable and can be ejected from the system entirely. 

In Fig.~\ref{fig:collisionage}, we show the ages at which the two stars in the system collide and merge as a function of initial orbital separation.
The four cases differ in the initial rotation rates of the two stars.
We exclude all simulations in which the two stars collided almost instantly (i.e. within a few thousand years), and all simulations in which the two stars did not collide.
In all cases, the age at which the two stars collide is higher for larger initial orbital separations.
The exact collison age depends sensitively on the initial rotation rates of the two stars.
When both stars start as slow rotators, the stars collide much earlier than they do when both stars start as fast rotators.
This is due to the initial evolution of the orbital separation as described above.

\section{XUV evolution in tight binary systems} \label{sect:XUVevo}

In this section, we describe the evolution of the X-ray and EUV emission from stars in tight binary systems and how it differs from single star systems.
Our approach is based on the approach of \citet{Tu15}.
As in \citet{Wright11}, we assume
\begin{equation} \label{eqn:wrightlaw}
R_\text{X} = \left \{
\begin{array}{ll}
R_{\text{X},\text{sat}}, & \text{if } Ro \le Ro_{\text{sat}},\\
C Ro^\beta, & \text{if } Ro \ge Ro_{\text{sat}},\\
\end{array} \right.
\end{equation}
where \mbox{$R_\text{X} = L_\text{X} / L_\text{bol}$}, $Ro_{\text{sat}}$ is the saturation Rossby number, $R_{\text{X},\text{sat}}$ is the saturation value of $R_\text{X}$, and $C$ and $\beta$ determine the X-ray relation in the unsaturated regime.
As in \citet{Wright11}, we use \mbox{$R_{\text{X},\text{sat}} = 10^{-3.13}$}, \mbox{$\beta = -2.7$}, and \mbox{$Ro_\text{sat} = 0.13$}.
We calculate the stellar EUV luminosity from the X-ray luminosity using \mbox{$\log_{10} L_\mathrm{EUV} = 4.8 + 0.86 \log_{10} L_\mathrm{X}$} (\citealt{2011A&A...532A...6S}).
The XUV luminosity is simply the sum of $L_\mathrm{X}$ and $L_\mathrm{EUV}$.
For the convective turnover times, we use the same values as we use in the rotation model.
As in \citet{Tu15}, we normalise these values by a constant factor to make them consistent with the convective turnover times used by \citet{Wright11} at the age of the Sun.

We show in Fig.~\ref{fig:binaryradiationtracks} the XUV evolutionary tracks for solar mass single stars with initially slow and fast rotation rates (see the dotted lines in all panels).
At ages younger than $\sim$10~Myr, both stars are in the saturated regime regardless of their rotation rates because they have long convective turnover times, and therefore small Rossby numbers.
At about $\sim$10~Myr, the slow rotator comes out of saturation, due to the decreasing convective turnover times, and its XUV emission drops rapidly by almost an order of magnitude. 
The rapid rotator remains saturated until an age of $\sim$350~Myr.
After dropping out of saturation, both stars slowly decay to the activity level of the current Sun.  
These tracks are described in more detail in \citet{Tu15}.

In Fig.~\ref{fig:binaryrotationtracks}, we show rotational evolution tracks for tight binary star systems with different initial orbital separations.
Each system is composed of two solar mass stars with one initially slow rotator and one initially rapid rotator.
At long initial separations of 0.3~AU or greater, the tidal interactions have only negligible effects on the XUV evolution tracks.
For initial separations of $\sim$0.2~AU, the tidal interactions moderately influence the early radiative evolution, and become important after several Gyr when the tidal interactions stop the stars spinning down and instead causes their rotation rates to increase.
For smaller initial orbital separations, tidal interactions are important at all evolutionary times.
For the initially slowly rotating star, the tidal interactions increase the surface rotation rate at all evolutionary stages and therefore increase the XUV emission.
For the initially rapidly rotating star, the tidal interactions decrease the surface rotation rate at young ages but also stop the star from spinning down at later ages, leading to decreased activity at young ages and increased activity when the star is older.

It is interesting to consider the total XUV energy emitted during the lifetimes of the two stars. 
In Fig.~\ref{fig:binarytotalradiation}, we show separately the total $L_\mathrm{XUV}$ of the binary system integrated until 1~Gyr and over the entire lifetime for all four cases as a function of initial orbital separation.
We also plot the quantities separately as functions of the initial orbital separation and the orbital separation at 5~Myr in order to show the influence of the initial orbital evolution described in the previous section.
As can be seen in Fig.~\ref{fig:binarytotalradiation}, at small orbital separations, the differences between the cases is entirely a result of this initial orbital evolution; when this effect is removed, all systems emit the same total amount of XUV radiation for a given orbital separation, regardless of the initial rotation rates of the two stars.
We find that in almost all cases, tidal interactions increase the total emitted XUV energy relative to the amount that would be emitted by the same single stars.
In the most extreme cases, this is an increase by a factor of $\sim$50.
In a small number of cases, we find that the total XUV emission is reduced by a factor of a few by tidal interactions; this is for initial orbital separations of $\sim$0.13~AU for Case~1 and Case~3, which both contain initially fast rotators.

In the absence of the tidal synchronisation torque, the total power emitted depends sensitively on the initial rotation rates of the two stars.
With orbital separations greater than $\sim$0.2~AU, tidal interactions have almost no effect on the total energy emitted. 
At smaller separations, the total energy emitted depends sensitively on the initial orbital separation.
For all cases, the trends are similar: decreasing the initial separation increases the total energy because the stars are made to rotate rapidly for a long time, until a certain orbital separation at which point decreasing the orbital separation further leads to a decrease in the total energy emitted.
This decrease is caused by the decay in the orbits of the binary systems causing them to collide and merge, and the fact that lower initial orbital separations lead to this collision happening earlier.
The decrease is likely reduced to some extent if we also included the radiation emitted by the single star that is formed after the two stars merge. 

The trends are slightly different when considering only the energy emitted in the first Gyr.
In this time period, the decrease in emitted energy at small separations is not present, simply because it takes usually more than 1~Gyr for the two stars to collide. 
However, the trend of energy emitted as a function of orbital separation is very different with decreasing orbital separation leading to significant reductions in the XUV energy emitted starting at separations of $\sim$0.25~AU, and then this trend reversing at $\sim$0.13~AU.
In most cases, this reduction is more than compensated for at later ages because the stars are kept rapidly rotating for a long time, so such stars emit overall much more energy.

In Fig.~\ref{fig:binarysaturationage}, we show the ages at which stars drop out of saturation as a function of initial orbital separation for all four cases.
The general trend is that at small orbital separations, rapidly rotating stars fall out of saturation earlier due to the additional spin-down caused by tidal interactions, and slowly rotating stars remain saturated longer due to the additional spin-up.
Systems with very small initial orbital separations (for example our 0.03~AU case shown in Fig.~\ref{fig:binaryradiationtracks}) remain saturated for their entire lifetimes.
It is even possible in some cases for tidally locked stars to drop out of saturation at a young age, and then to become saturated again later due to the spin-up caused by their decreasing orbital separations.

\section{Planetary atmosphere erosion} \label{sect:atmosphereevo}

\begin{figure*}
\centering
\includegraphics[trim = 0mm 0mm 0mm 0mm, clip=true,width=0.45\textwidth]{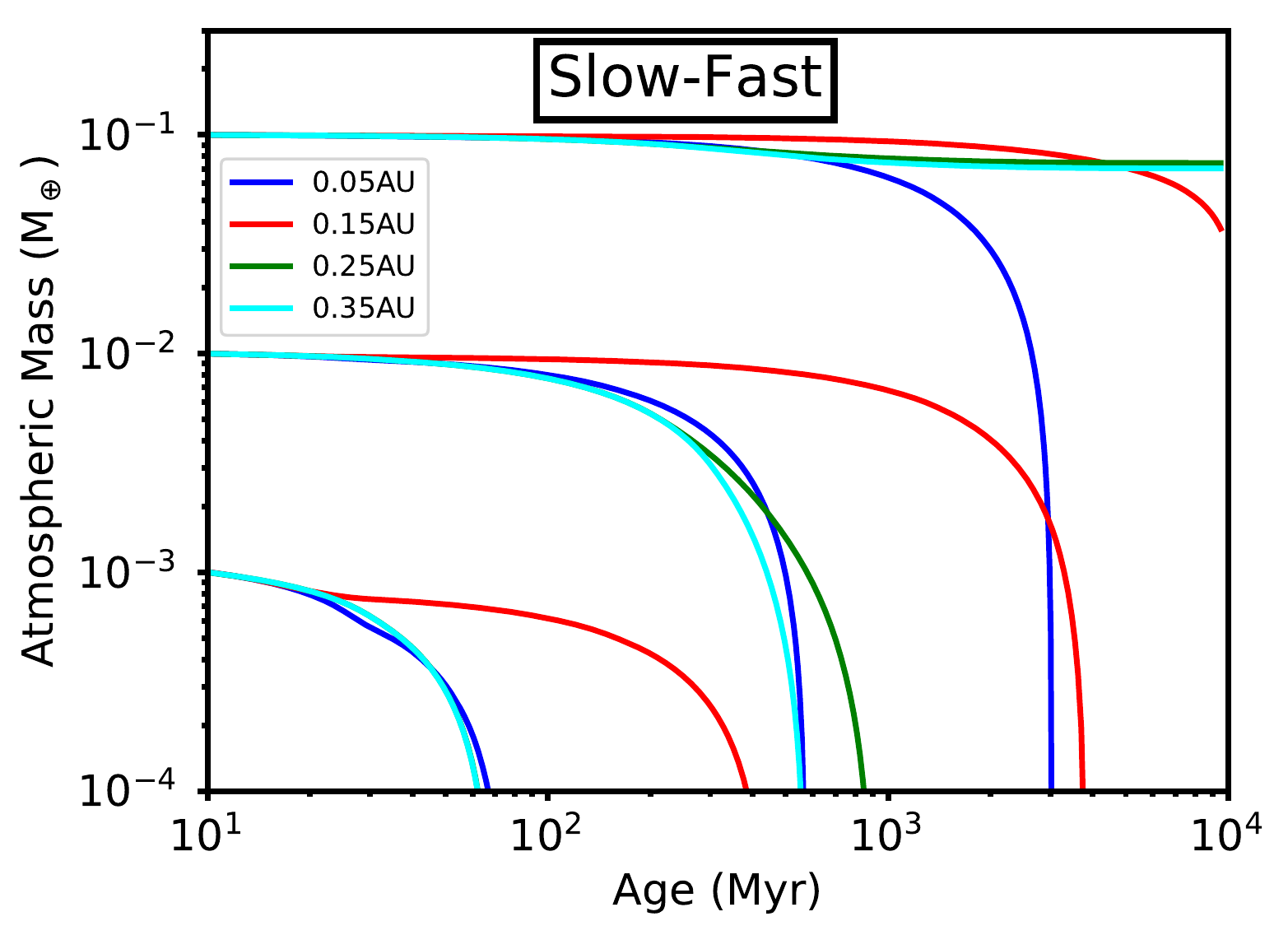}
\includegraphics[trim = 0mm 0mm 0mm 0mm, clip=true,width=0.45\textwidth]{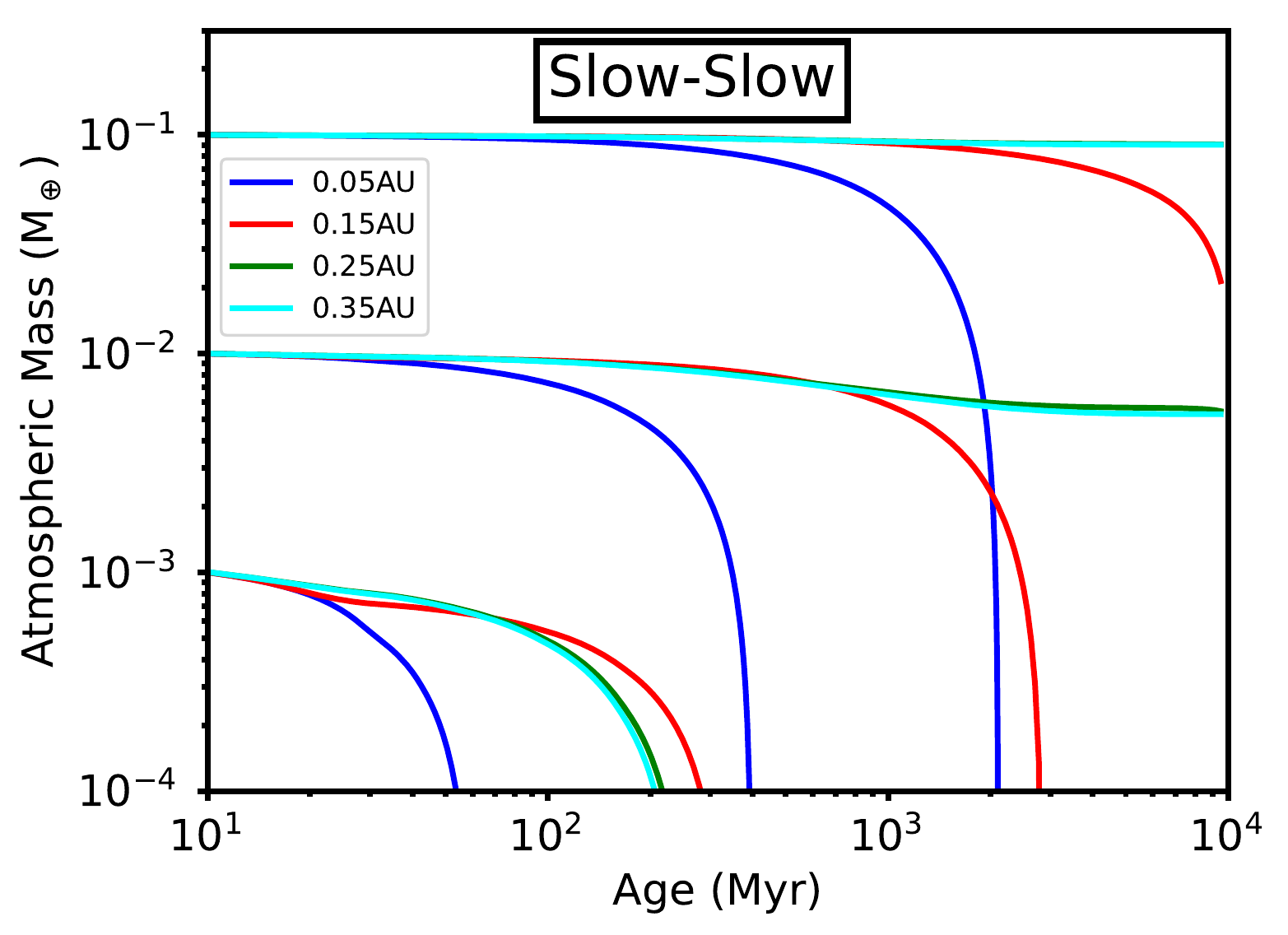}
\includegraphics[trim = 0mm 0mm 0mm 0mm, clip=true,width=0.45\textwidth]{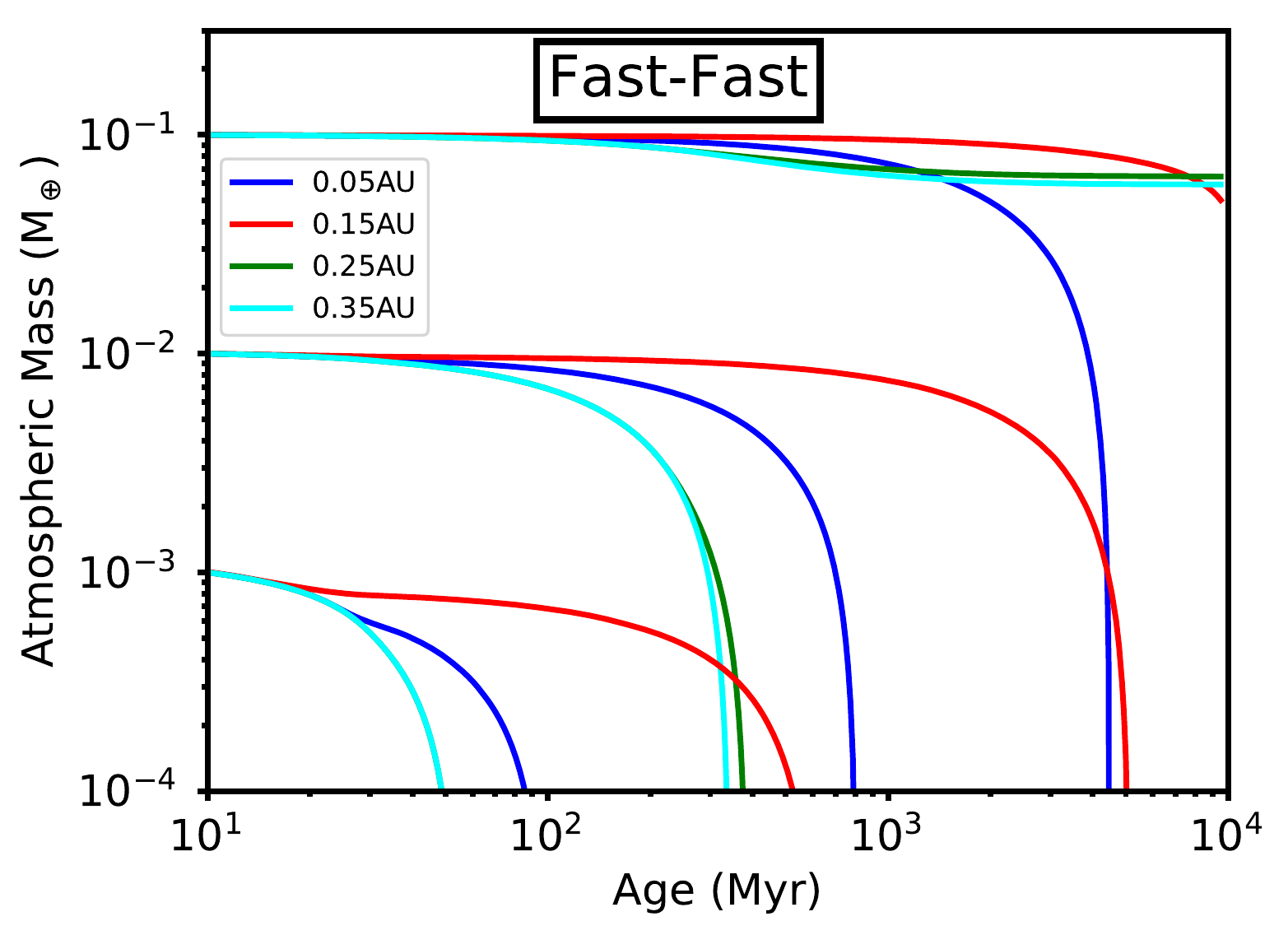}
\includegraphics[trim = 0mm 0mm 0mm 0mm, clip=true,width=0.45\textwidth]{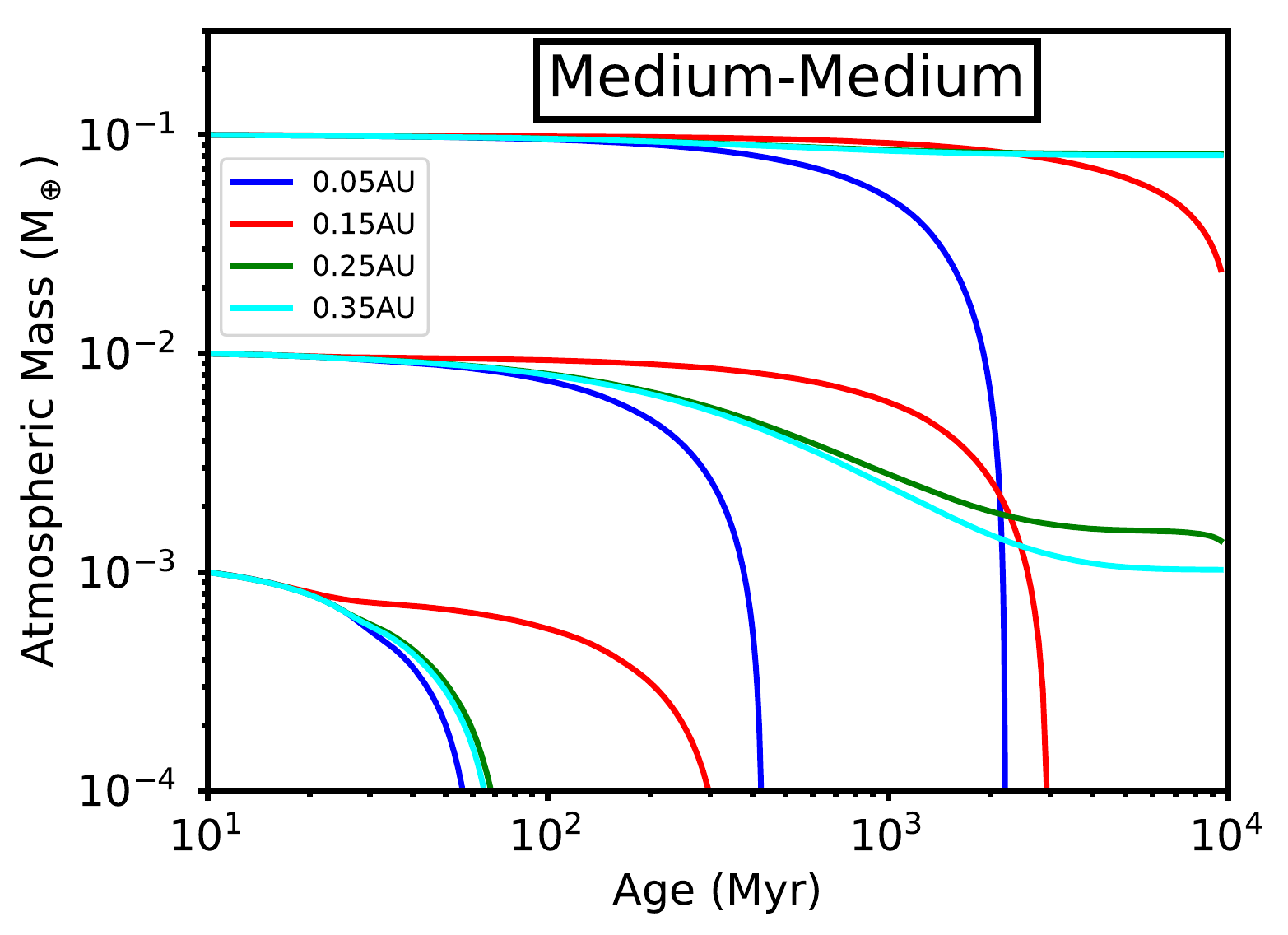}
\caption{
Evolutionary tracks of the atmospheric mass for Earth mass planets with hydrogen atmospheres orbiting tidally interacting solar mass stars.
Each line shows a different case, with the three sets of lines on each panel showing cases with different initial (10~Myr) atmospheric masses, the different colours representing different initial orbital separations of the two stars, and the two panels showing the tracks with different initial rotation rates of the two stars.
The left panel shows our Case~2, where both stars start as slow rotators, and the right panel shows our Case~3, where both stars start as fast rotators.
}
 \label{fig:MatTracks}
\end{figure*}

\begin{figure}
\centering
\includegraphics[trim = 0mm 0mm 0mm 0mm, clip=true,width=0.45\textwidth]{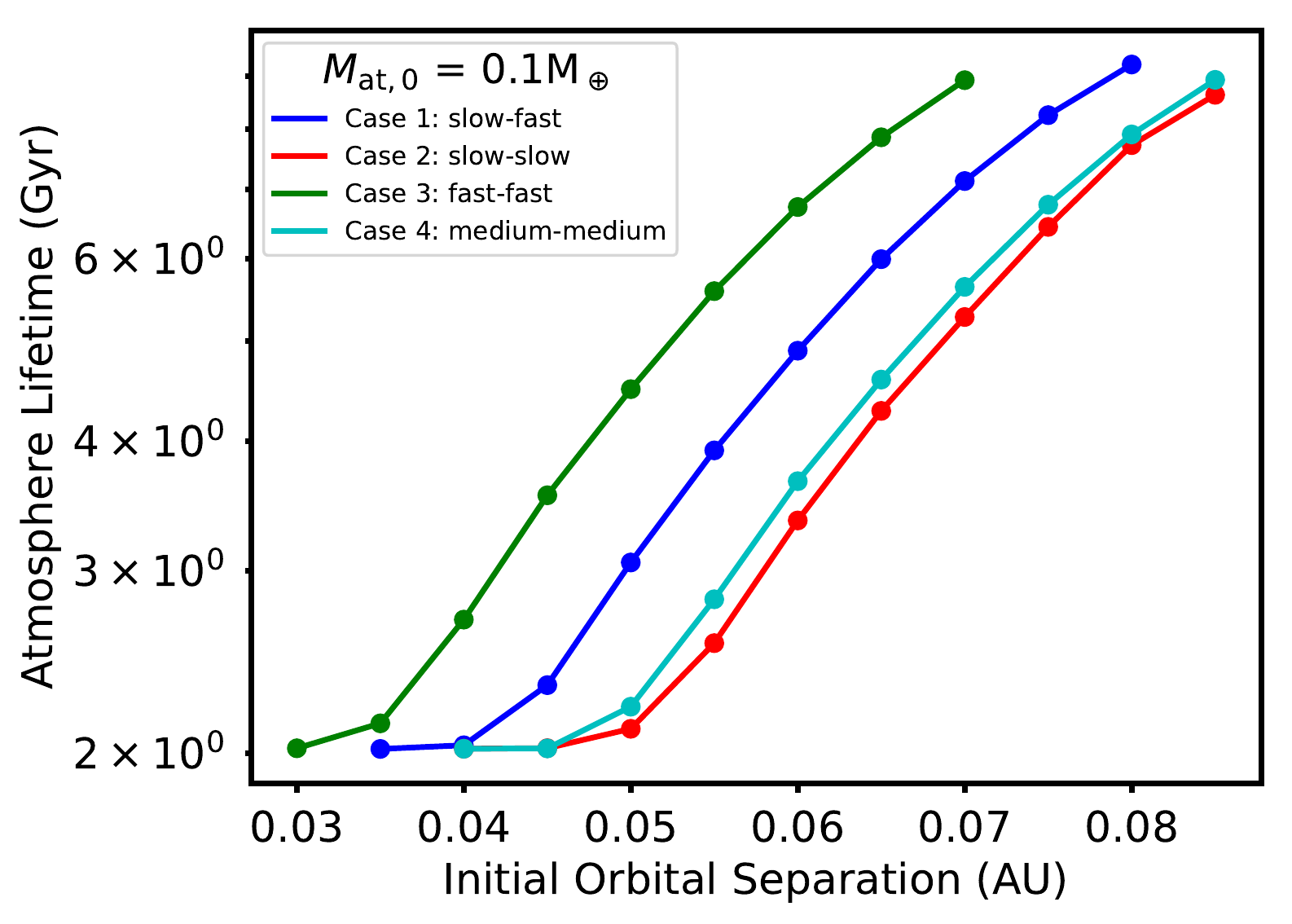}
\includegraphics[trim = 0mm 0mm 0mm 0mm, clip=true,width=0.45\textwidth]{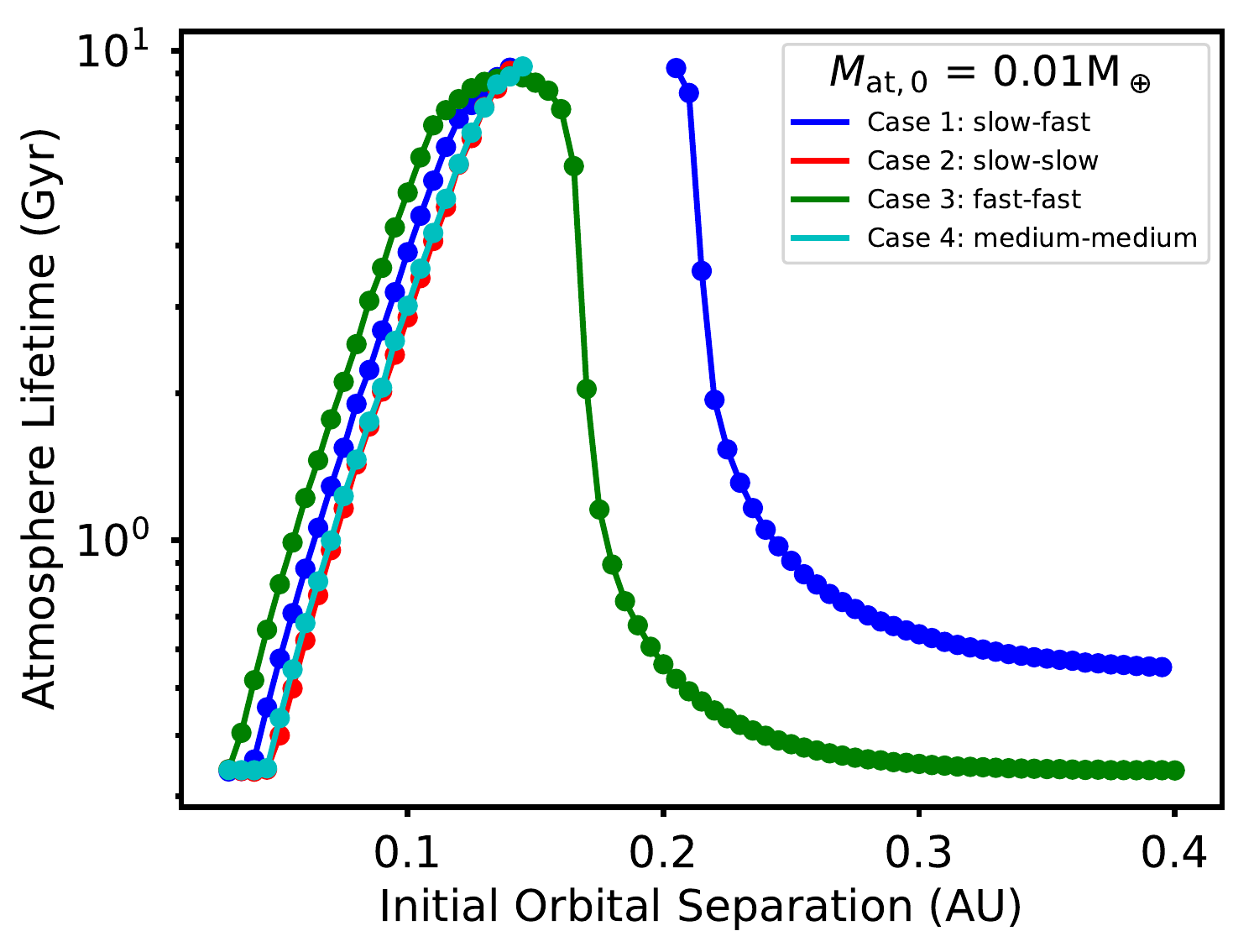}
\includegraphics[trim = 0mm 0mm 0mm 0mm, clip=true,width=0.45\textwidth]{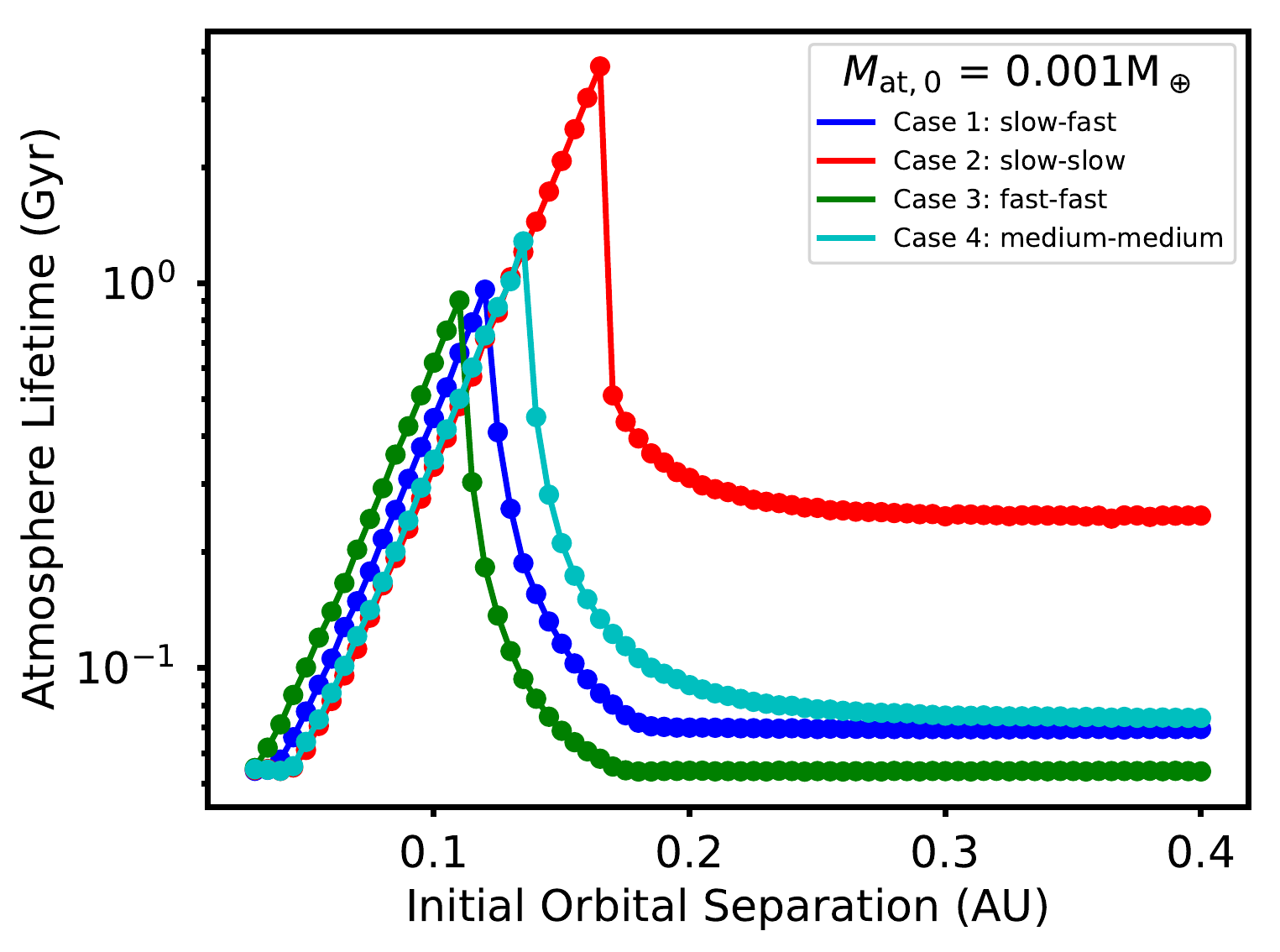}
\caption{
Amount of time taken to remove the entire planetary atmosphere as a function of initial stellar orbital separation.
The three panels show different initial atmospheric masses, as indicated in the top-left corners, and the different lines give the results assuming different initial rotation rates for the two stars.
Only simulations in which the atmospheres were removed during our evolutionary simulations are included. 
}
 \label{fig:atLifetime}
\end{figure}

Planetary atmospheres form in several ways.
Planets that form quickly can pick up significant envelopes of light (mostly hydrogen) gas from the circumstellar gas disk (\mbox{\citealt{1979E&PSL..43...22H}}), which are often called `primordial atmospheres'.
For such protoatmospheres to form, the cores must form to a significant mass (\mbox{$\gtrsim 0.1$~M$_\oplus$})  within a few Myr while the gas disk is present.
Later on, when the planet cools and its surface solidifies, a secondary atmosphere of heavier gases, such as N$_2$ and CO$_2$, can form (\mbox{\citealt{2008E&PSL.271..181E}}; \mbox{\citealt{Noack14}}). 
In both cases, removing the atmosphere can be difficult.
For secondary atmospheres, this is mostly due to the high molecular masses of the gases in the atmosphere and, in the case of CO$_2$ dominated atmospheres, the strong cooling of the thermosphere by infrared radiation. 
For hydrogen dominated atmospheres, this is mostly due to the huge amounts of gas that can be captured during the disk phase.
Depending on its mass and how long it spends in the disk, a core can quite easily capture an atmosphere that is a few percent of its own mass (\citealt{Stoekl16}). 
Observational evidence that atmospheric losses can have a significant effect on primoridal atmospheres has been identified in radii measurements of exoplanets by \emph{Kepler} (\citealt{OwenWu17}), and active atmospheric losses have also likely been observed directly (\mbox{\citealt{2008A&A...483..933E}}).

The detection of several low-mass and high radius planets suggests that not only do terrestrial planets with massive hydrogen envelopes exist, but are in fact very common (\mbox{\citealt{2011Natur.470...53L}}; \mbox{\citealt{2015ApJ...801...41R}}).
The formation and loss of these atmospheres in the first 100~Myr was modelled by \mbox{\citet{2014MNRAS.439.3225L}} and \citet{OwenMohanty16} who found that the final outcome (i.e. whether or not a planet kept its primordial atmosphere) was primarily determined by the mass of the core. 
Cores less massive than the Earth tended to accumulate much less atmosphere during the disk phase and lose it quickly, whereas cores more massive than the Earth tended to accumulate large atmospheres that they were not able to lose at any point. 
\citet{Johnstone15apj} showed that the evolution of a planet's primordial atmosphere can depend on the initial rotation rate of the host star, in single star systems, due to the different XUV evolution pathways for stars with different rotation rates.

In tight binary systems, when the energy emitted by the two stars over the lifetime of the system is enhanced by tidal-interactions, much more atmospheric gas should be lost than we would expect in single star systems, and the amount that can be lost depends on the orbital separations of the two stars.
This can be understood by considering the energy-limited formula for hydrodynamic escape, which gives \mbox{$\dot{M}_\mathrm{at} \propto F_\mathrm{XUV}$} if all the planetary parameters are constant (\citealt{Erkaev07}).
This implies that the total atmospheric mass loss is approximately proportional to the total energy emitted by the stars, which we discuss in the previous section.
Hydrodynamic models of atmospheric expansion and escape have shown that the mass loss rate is not exactly proportional to the input XUV flux (e.g. see Fig.~1 of \citealt{Johnstone15apj}).

In this section, we concentrate on atmospheres composed primarily of hydrogen and we consider only hydrodynamic escape of atmospheres.
Since strong XUV-driven hydrodynamic losses are also expected to take place on terrestrial planets with Earth-like atmospheres (Johnstone et al. 2019), we will study atmospheres with different compositions in future work.

\subsection{Atmospheric loss model} \label{sect:atmodel}

The mechanism for mass loss that we consider in this paper is hydrodynamic flow driven by stellar XUV heating of the upper atmosphere (\mbox{\citealt{1981Icar...48..150W}}; \mbox{\citealt{2005ApJ...621.1049T}}). 
For hydrogen dominated atmospheres, this is likely the dominant mechanism (\mbox{\citealt{Erkaev13}}; \mbox{\citealt{2013AsBio..13.1030K}}).
Due to the large absorption cross sections of atmospheric gas in X-ray and EUV wavelengths, XUV radiation is absorbed high up in the atmosphere.
At the base of the thermosphere, the gas has almost no bulk motion in the radial direction and a relatively low temperature, approximately equal to the planet's effective temperature. 
Higher in the thermosphere, the absorption of XUV radiation rises the temperature dramatically to accelerate away from the planet to speeds of several km~s$^{-1}$ or more.
The combination of the radial expansion and the acceleration means that the wind density decreases rapidly with altitude, and at the exobase, the gas becomes collisionless.
If the speed has reached the escape velocity below the exobase, then the planet will be losing material in the form of transonic hydrodynamic outflow. 
In such a wind, the accelerating wind reaches the escape velocity at the sonic point.

Our atmosphere loss model, which is described in \mbox{\citet{Johnstone15apj}} and is similar to the model of \citet{Erkaev13}, uses the Versatile Advection Code (\mbox{\citealt{1996ApL&C..34..245T}}) to simulate the planetary thermosphere in a 1D spherically symmetric geometry.
This is a simplified atmosphere model that neglects many details of the thermal and chemical processes acting in the atmosphere.
Our assumption of 1D spherical symmetry might lead to some overestimation of the loss rates (\citealt{StoneProga09}).
In future work, we will study this problem using  more sophisticated upper-atmosphere models, such as the model recently developed model \citet{Johnstone18}. 
To calculate the time evolution of an atmosphere, \mbox{\citet{Johnstone15apj}} produced a grid of thermosphere models with different input parameters.
From their grid, they found that the atmospheric mass loss rate, $\dot{M}_\text{at}$, can be expressed as
\begin{equation} \label{eqn:Mdoteqn}
\dot{M}_\text{at} = a m_\text{H} M_\text{pl}^b z_0^c \left( \log F_\text{XUV} \right)^{ g \left( M_\text{pl} , z_0 \right) },
\end{equation}
where
\begin{equation}
g \left( M_\text{pl} , z_0 \right) = d M_\text{pl}^e z_0^f,
\end{equation}
$M_\text{pl}$ is the planetary mass, $m_\text{H}$ is the mass of a hydrogen atom, and \mbox{$z_0 = R_0 - R_\text{core}$} is the altitude of the base of the simulation. 
With $M_\text{pl}$, $z_0$, and $F_\text{XUV}$ in units of $M_\oplus$, $R_\oplus$, and erg~s$^{-1}$~cm$^{-2}$, they found that \mbox{$a = 1.858 \times 10^{31}$}, \mbox{$b = -1.526$}, \mbox{$c = 0.464$}, \mbox{$d = 4.093$}, \mbox{$e = 0.249$}, and \mbox{$f = -0.022$}.
To calculate $z_0$ as a function of $M_\text{pl}$ and $M_\text{at}$, they ran a grid of models for the structure of the lower atmosphere of a hydrogen envelope using the initial model integrator of the TAPIR-Code (\mbox{\citealt{2015A&A...576A..87S}}) to solve the hydrostatic structure equations.
They derived the following expression
\begin{equation} \label{eqn:R0scaling}
\log \left( \frac{R_0}{R_\text{core}} \right) = \left( 2.5 f_\text{at}^{0.4} + 0.1 \right) \left( \frac{M_\text{pl}}{M_\oplus} \right)^{-0.7},
\end{equation}
where $R_\text{core}$ is the radius of the solid core and \mbox{$f_\text{at} = M_\text{at} / M_\text{pl}$} is the envelope mass fraction.
Combining these equations allows us to integrate the atmospheric mass in time assuming arbitrary planetary masses, intial atmospheric masses, and evolutionary tracks for the stellar XUV luminosity, taking into account the shrinking of the atmosphere as its mass decreases.

\subsection{Results: atmospheric evolution for an Earth mass planet}

Since we are interested primarily in habitable zone planets, we assume the planet is on a circular orbit with a distance of 1.48~AU from the centre of mass of the system.
Assuming the current solar luminosity for both stars and an albedo of 0.3, which is approximately that of both Neptune and Earth, this orbital distance gives an effective temperature averaged over an orbit of 250~K. 
The orbital variations in the effective temperature are never more than a few K in all cases considered in this paper.  
In reality, even if the orbits start out circular, gravitational pertubations by the two stars will cause the orbits to periodically gain and lose eccentricity (\mbox{\citealt{2013ApJ...777..166H}}).
However, given the parameters of our system, the eccentricity gained will likely be negligible (e.g. see Fig.~7 of \mbox{\citealt{2015A&A...577A.122J}}).
Also, the variations of $F_\text{XUV}$ over an entire orbit are very small given that the orbital separations that we assume for the stars are significantly smaller than the orbital radius of the planet. 

To demonstrate how the orbital distance between the two stars in tidally locked binary systems influences the atmospheric structure and the mass loss rate, we use the results from the previous sections to run a series of atmospheric evolution models.
In these models, we evolve the planetary atmospheres assuming the XUV evolution tracks discussed in Section~\ref{sect:XUVevo}.
We start all of our atmospheric evolution models at 10~Myr and evolve the atmospheric mass until the end of the main-sequence lifetimes of the stars, or until the two stars merge, or until the atmosphere has been completely lost.
Since we assume that the planet's mass is 1~M$_\oplus$, the only planetary parameter that we vary is the initial mass of the atmosphere.
We assume initial masses of 0.1, 0.01, and 0.001~M$_\oplus$.
The other three input parameters are the initial rotation rates of the two stars and the initial orbital separation.
Evolutionary tracks for the atmosphere masses are shown in Fig.~\ref{fig:MatTracks}.
In Fig.~\ref{fig:atLifetime}, we show the ages at which the entire atmospheres of the planets have been removed as a function of orbital separation for all four cases, and for all three assumed initial atmospheric masses.
Models in which the atmospheres remained at the end of the simulation are not included.

In all cases, the orbital separation is very important for determining how long it takes for an atmosphere to be completely lost. 
Consider first the case in which both stars start as slow rotators and the initial atmospheric masses are 0.01~M$_\oplus$ (i.e. 1\% of the planet's mass).
For a separation of 0.25~AU or more the atmosphere survives until the end of the main-sequence lifetimes of the two stars, and for initial separations of 0.15 and 0.05~AU, the entire atmospheres are lost in 3~Gyr and 400~Myr respectively.  
In the models that start with much less massive atmospheres, the atmospheres are always lost quite rapidly, but in times that are anyway sensitively dependent on orbital separation.
For different combinations of initial rotation rates, the results are also mostly similar, except for the cases with initial atmospheric masses are 0.01~M$_\oplus$ and initial orbital separations of 0.25 and 0.35~AU; in these cases, the atmospheres are list only when at least one of the two stars start as a fast rotator. 

The atmospheric lifetimes shown in Fig.~\ref{fig:atLifetime} show complex dependences on initial orbital separation. 
The cases with initial atmospheric masses of 0.1~M$\_odot$ are the simplest, with most orbital separations leading to the atmospheres surviving entirely.
With initial separations below approximately 0.08~AU, the total emitted XUV flux becomes high enough that the atmosphere can be entirely removed within the lifetime of the system, and how long the atmosphere lasts depends sensitively on the separation. 
This is significant since it would otherwise be impossible for an Earth mass planet to lose an 0.1~M$_\odot$ hydrogen atmosphere when orbiting at 1~AU around a single solar mass star, even if the star starts its life as a rapid rotator (\citealt{2015ApJ...815L..12J}).

With an initial atmospheric masses of 0.01~M$\_odot$, it is possible for the atmospheres to be destroyed within the system's lifetimes with larger binary separations if one of the stars is born as a rapid rotator. 
Interestingly, going to smaller orbital separations, the atmospheric lifetimes in these cases actually increase due to the reduction in the total XUV energy emitted, as can be seen in Fig.~\ref{fig:binarytotalradiation}.
This result is consistent with the suggestion of \citet{2015IJAsB..14..391M} that tidal interactions can increase the abilities of planets to retain their atmospheres. 
At orbital separations below approximately 0.15~AU, the trend reverses and decreasing orbital separations always leads to atmospheres being lose more rapidly.
The trends for the cases with initial atmospheric masses of 0.001~M$\_odot$ are similar, with significantly shorter atmospheric lifetimes.

\section{Discussion} \label{sect:conclusions}

In this paper, we study the evolution of stellar rotation, orbital separation, stellar XUV emission, and planetary atmospheres in binary star systems consisting of two tidally interacting solar mass stars and a habitable zone Earth mass planet.
We show that tidal interactions significantly influence rotational evolution when the orbital separations are less than $\sim$0.1~AU, and can even have a large effect on the later rotational spin-down at larger separations.
The combination of stellar wind angular momentum removal and tidal interactions can also cause significant orbital evolution, in many cases leading to the two stars merging.
In many of our models, significant orbital evolution takes place as the winds remove angular momentum from the system.
This could influence the orbital motion of the planet, especially if a resonance between the stellar and planetary orbits occurs. 
To which extent the planetary orbit is perturbed will be subject of a further investigation.

The tidal interactions between the two stars can have a strong influence on the amount of XUV energy that the two stars emit over their lifetimes.
The genera when the orbital separations are less than $\sim$0.12~AU, systems with smaller orbital separations emit more XUV energy than systems with larger separations, though at very short separations, this trend is reversed due to the earlier merging of the two stars.
This increase in the XUV emission can significantly increase atmospheric losses for planets orbiting the two stars.
We show that Earth mass planets with hydrogen atmospheres can lose significantly more atmospheric gas in these systems than they can in the habitable zone of a single star system, though how important the tidal interactions are depends on the specific parameters of the system.

A quite striking result of these models is that when the orbital separation is very small, even an atmosphere with an initial mass of 0.1~M$_\oplus$ can be removed.
This is interesting, not because an atmosphere with such a large mass is likely to form around an Earth mass planet, but because such a large atmosphere cannot be removed when the stars are evolving as single stars unless the planet is much closer to the star.
For planets in the habitable zones of tidally interacting binary systems, much larger amounts of atmospheric gas can be removed than is possible in the habitable zones in single star systems.
It is unclear if this would make the formation of habitable planetary environments in tight binary systems more likely because it is easier for planets to lose large undesirable hydrogen envelopes and other undesirable types of atmospheres (e.g. thick Venus-like CO$_2$ atmospheres), or less likely because the enhanced atmospheric removal could also strip away desirable secondary atmospheres (e.g. Earth-like N$_2$ atmospheres). 
Further theoretical work considering a more diverse set of atmosphere compositions will be necessary to answer this question.

When the two stars have initial orbital separations between approximately 0.12~AU and 0.2~AU, it is possible for tidal interactions to cause the systems to emit less XUV energy over their lifetimes than they would in single star systems. 
This is only the case is the stars start their lives are relatively rapid rotators since tidal interactions would cause the stars to spin down more rapidly than they otherwise would. 
Since most stars are not born as rapid rotators, this effect is likely only important in a minority of tight binary systems. 
We see from our atmospheric evolution models that this effect can protect an atmosphere from being eroded in some cases.
This is consistent with the ideas of \citet{2013ApJ...774L..26M} and \citet{2015IJAsB..14..391M} that tidal interactions in tight binary systems can protect atmospheres. 
Similar to the discussion above, it is however unclear if this will make the formation of a habitable planet more likely, as they suggest, or less likely. 

In some cases, the orbital evolution causes the two stars to collide and eventually to merge.
It is interesting to consider what happens in such systems when this happens.
The result of such a merger is a single star with a mass equal to the sum of the masses of the two merged stars.
The bolometric stellar luminosity depends strongly on stellar mass; for solar mass stars, this is approximately given by \mbox{$L_\mathrm{bol} \propto M_\star^4$}, which means that the merging of two solar mass stars leads to a single star that is a factor of eight brighter than the two stars were previously.
This would cause the inner habitable zone boundary to be shifted from 1.34 to 2.87~AU and the outer habitable zone boundary to be shifted from 2.37 to 4.95~AU (calculated using the method of \citealt{Kopparapu2014}).
When this happens, previously frozen planets can become habitable.
However, it is unclear whether or not planets in such systems could become habitable given that such mergers could be accompanied by a large outburst as the orbital energy is released. 
This was seen in 2008 in the case of V1309 Sco, though the outbust lasted only approximately a year (\citealt{Tylenda11}). 
Further research is needed to study what the effects of such an outburst on planetary habitability.


\section{Acknowledgments} 

This study was carried out with the support by the Austrian Science Fund (FWF) project S11601-N16 ``Pathways to Habitability: From Disk to Active Stars, Planets and Life'' and the related subprojects S11604-N16, and S11608-N16.

\appendix

\section{Rotation model for short synchronisation timescales} \label{appendix:modifiedmodel}

The basic assumption that we make when the synchronisation timescales are extremely short is that the stars are perfectly synchronised, such that
\begin{equation}
\frac{d \Omega_\mathrm{env,1}}{dt} = \frac{d \Omega_\mathrm{env,2}}{dt} = \frac{d \Omega_\mathrm{orb}}{dt} = \frac{d \Omega}{dt},
\end{equation}
where $d \Omega/dt$ refers to all three rates of change.
It is not necessary to assume additionally that \mbox{$\Omega_\mathrm{env,1}=\Omega_\mathrm{env,2}=\Omega_\mathrm{orb}$} since our model only uses the above assumption when the three rotation rates are anyway equal to within 0.1\%.
As in Eqn.~\ref{eqn:dOmegaEnvdt}, the equation for the torque acting on the envelope of a star is \mbox{$d J_\mathrm{env}/dt = \tau_\mathrm{w} + \tau_\mathrm{ce} + \tau_\mathrm{cg} + \tau_\mathrm{dl} + \tau_\mathrm{ts}$}.
Rearanging this for $\tau_\mathrm{ts}$ and inputting it into Eqn.~\ref{eqn:dJorbdt} for both stars gives
\begin{equation} \label{eqn:angmomsync}
\begin{split}
\frac{d}{dt} & \left( J_\mathrm{env,1} + J_\mathrm{env,2} + J_\mathrm{orb} \right) 
= (\tau_\mathrm{w,1}+\tau_\mathrm{w,2})\\
& + (\tau_\mathrm{ce,1}+\tau_\mathrm{ce,2}) + (\tau_\mathrm{cg,1}+\tau_\mathrm{cg,2}) + (\tau_\mathrm{dl,1}+\tau_\mathrm{dl,2}),
\end{split}
\end{equation}
where the sub-scripts 1 and 2 refer to the quantities for the two stars. 
Combining Eqn.~\ref{eqn:Jorb} with \mbox{$a_\mathrm{orb}^3 = G (M_{\star,1} + M_{\star,2}) / \Omega_\mathrm{orb}^2$} gives
\begin{equation}
\frac{dJ_\mathrm{orb}}{dt} = - \frac{ G^{\frac{2}{3}} M_{\star,1} M_{\star,2} }{ 3 \left(M_{\star,1} + M_{\star,2} \right)^{\frac{1}{3}} \Omega_\mathrm{orb}^{\frac{4}{3}} } \frac{d \Omega_\mathrm{orb}}{dt} .
\end{equation}
Inserting this and \mbox{$J_\mathrm{env} = I_\mathrm{env} \Omega_\mathrm{env}$} into Eqn.~\ref{eqn:angmomsync} gives
\begin{multline}
\left[ I_\mathrm{env,1} + I_\mathrm{env,2} - \frac{G^{\frac{2}{3}} M_{\star,1} M_{\star,2} }{3 \left(M_{\star,1} + M_{\star,2} \right)^{\frac{1}{3}} \Omega_\mathrm{orb}^{\frac{4}{3}}} \right] \frac{d\Omega}{dt} =  \\
\sum_{i=1}^{2} \left[ \tau_\mathrm{w,i} + \tau_\mathrm{ce,i} + \tau_\mathrm{cg,i} + \tau_\mathrm{dl,i} - \Omega_\mathrm{env,i} \frac{d I_\mathrm{env,i}}{dt}  \right],
\end{multline}
where the sum is over both stars. 
The $d\Omega/dt$ from this equation gives us $d\Omega_\mathrm{env}/dt$ for both stars and is used to calculate $d a_\mathrm{orb}/dt$ using
\begin{equation}
\frac{d a_\mathrm{orb}}{dt} = - \frac{2}{3} G^{\frac{1}{3}} \left( M_{\star,1} + M_{\star,2} \right)^{\frac{1}{3}} \Omega_\mathrm{orb}^{-\frac{5}{3}} \frac{d \Omega}{dt}.
\end{equation}
The evolution equations for the core rotation rates remain unchanged from the original model.

\bibliographystyle{aa}
\bibliography{mybib}

\end{document}